\documentclass[a4paper,11pt]{article}
\pdfoutput=1 

\usepackage{jheppub} 
\usepackage[T1]{fontenc} 
\usepackage{latexsym}
\usepackage{amsmath}
\usepackage{graphicx}
\usepackage{caption}
\usepackage{tabu}
\usepackage{amssymb}
\usepackage{amsmath}
\usepackage{amstext}
\usepackage{graphicx,epsfig}
\usepackage{epsfig}
\usepackage{verbatim} 
\usepackage{fancybox}
\usepackage{color}
\usepackage{ulem}
\usepackage{enumitem}
\usepackage{subfigure}
\usepackage{bbm}
\usepackage{parskip}
\usepackage{dsfont}
\usepackage[numbers,sort&compress]{natbib}
\usepackage{bm}
\usepackage[dvipsnames]{xcolor}
\usepackage{placeins}
\DeclareMathOperator{\Lvert}{\big\vert}
\DeclareMathOperator{\ud}{\mathrm{d}}
\DeclareMathOperator{\M}{\mathcal{M}}

\title{\boldmath Corrections to di-Higgs boson production  with light stops and modified Higgs couplings}

\author[a,b]{Peisi Huang,}
\author[c,e,f]{Aniket Joglekar,}
\author[c,g]{Min Li,}
\author[c,d,h]{ Carlos E.M. Wagner}

\affiliation[a]{Mitchell Institute for Fundamental Physics and Astronomy, Department of Physics  and Astronomy, Texas A$\&$M University,
College Station, TX 77843,}
\affiliation[b]{Department of Physics and Astronomy, University of Nebraska-Lincoln, Lincoln, NE, 68588}
\affiliation[c]{Enrico Fermi Institute, University of Chicago, Chicago, IL 60637, }
\affiliation[d]{Kavli Institute for Cosmological Physics, University of Chicago, Chicago, IL 60637,}
\affiliation[e]{Amherst Center for Fundamental Interactions, Department of Physics,
University of Massachusetts Amherst, Amherst, MA 01003,}
\affiliation[f]{Department of Physics and Astronomy, University of California-Riverside,
900 University Ave., Riverside, CA 92521,}
\affiliation[g]{Department of Physics, University of Illinois at Urbana-Champaign, Urbana, Illinois 61801,}
\affiliation[h]{HEP Division, Argonne National Laboratory, 9700 Cass Ave., Argonne, IL 60439.}

\emailAdd{peisi.huang@unl.edu}
\emailAdd{aniket@ucr.edu}
\emailAdd{minl2@illinois.edu}
\emailAdd{cwagner@hep.anl.gov}

\abstract{The Higgs pair production in gluon fusion is a sensitive probe of beyond-Standard Model (BSM) phenomena and its detection is a major goal for the LHC and higher energy hadron collider experiments.  In this work we reanalyze the possible modifications of the Higgs pair production cross section within low energy supersymmetry models. We show that the supersymmetric  contributions to the Higgs pair production cross section are strongly correlated with the ones of the single Higgs production in the gluon fusion channel. Motivated by the analysis of ATLAS and CMS Higgs production data, we show that the scalar superpartners' contributions may lead to significant modification of the di-Higgs production rate and invariant mass distribution with respect to the SM predictions. 
We also analyze the combined effects on the di-Higgs production rate of a modification of the Higgs trilinear and top-quark Yukawa couplings in the presence of light stops. In particular, we show that due to the destructive interference of the triangle and box amplitude contributions to the di-Higgs production cross section, even a small modification of the top-quark Yukawa coupling can lead to a significant increase of the di-Higgs production rate.}

\preprint{EFI-17-23}

\begin{document} 
\maketitle
\flushbottom

\section{INTRODUCTION}
\label{sec:intro}

A scalar resonance with mass of approximately 125~GeV has been detected  at the run-I of the LHC~\cite{Aad:2012tfa, Chatrchyan:2012xdj}. Since its discovery, there has been a lot of effort in studying its properties.  In particular, the main production rates and decay modes at the LHC have been analyzed, leading to results that are close to the ones predicted for the Higgs boson in the SM. The accuracy of each of these measurements is low and hence at present a departure of the SM properties may only be obtained by a combined analysis of all production and decay channels. A recent combined analysis of the Higgs data collected at run I by ATLAS and CMS~\cite{Khachatryan:2016vau,Aad:2015zhl} in different production channels was used to determine the best fit to $\kappa_i = g_{hii}/g_{hii}^{SM}$, the ratio of the Higgs couplings with respect to the SM predicted values. All relevant ratios $\kappa_i$ are consistent with unity at the 2$\sigma$ level, although errors are still large and  moderate deviations of the Higgs couplings with respect to the SM values are possible. In fact, the best fit values of $\kappa_{i}$  present moderate deviations with respect to the  SM predictions, which allows  the presence of BSM effects in  Higgs physics.  

\begin{figure}[tbp]
	\centering
	\includegraphics[width=1\textwidth]{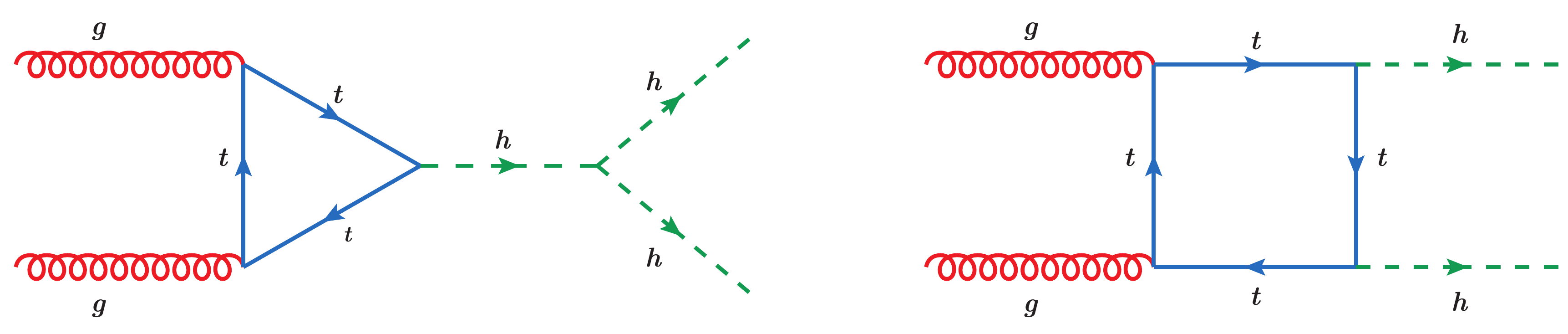}
	\caption{\label{gghh_fulp} SM Top-quark loop in Higgs pair production}
\end{figure}

The double Higgs production provides a probe for new physics. In the SM, at the leading order, the Higgs pair production process in gluon fusion, $gg\rightarrow hh$, receives contributions from two different quark-loop induced amplitudes, corresponding to a triangle  ($gg\rightarrow h^{*} \rightarrow hh$), and a box diagram, ($gg\rightarrow hh$), shown in  Fig.~\ref{gghh_fulp}, with top-quarks giving the main contribution. The amplitudes associated with the two diagrams interfere with each other destructively. When the di-Higgs invariant mass, $m_{hh}$, is below the threshold for the top quarks in the QCD loop to be produced on-shell, $m_{hh}\le 2 \ m_{t}$, the amplitudes associated with these two diagrams only contains real parts, and the destructive interference leads to an exact cancellation of the total one-loop amplitude at the di-Higgs production threshold $m_{hh} = 2 \ m_h$. The resulting cross section in the SM is small and the statistical significance of the Higgs pair production process becomes very low, making this process very sensitive to possible
deviations of these amplitudes from their SM values. 

The triangle and the box diagrams are both very sensitive to the Higgs couplings of the colored particles running in the QCD loop. This means that even a small deviation of the Higgs coupling to the top quark with respect to its SM value may lead to considerable impact on their contributions to the di-Higgs production. In addition, the triangle diagram, where a single off-shell Higgs is produced and transforms into a pair of Higgs bosons, is correlated to the diagram of single Higgs production via gluon fusion, $gg\rightarrow h$, and it is also proportional to the triple Higgs coupling, $\lambda_3$, which provides an important information in probing the Higgs potential. Moreover, the box and the triangle loop amplitudes become very sensitive to new heavy colored particles running in the QCD loop. Therefore, the di-Higgs production detection becomes a very promising channel in probing new physics, being sensitive to various kinds of new effects~\cite{Belyaev:1999mx,Batell:2015koa,Dolan:2012ac,Dawson:2015oha,Goertz:2013kp,Cao:2013si,Hespel:2014sla,Goertz:2014qta,Azatov:2015oxa,Cao:2015oaa,Cao:2016zob,Maltoni:2016yxb,Menon:2004wv,Grojean:2004xa,Noble:2007kk,Barger:2011vm,Chung:2012vg,Katz:2014bha,Curtin:2014jma,He:2015spf,Huang:2015tdv,Espinosa:2011ax,Profumo:2007wc,Profumo:2014opa,Kozaczuk:2015owa,Kotwal:2016tex,Morrissey:2012db,Chiang:2017nmu}. 

In this article, we shall analyze the possible modifications of the di-Higgs production rate within low energy supersymmetry models. These models allow for the presence of new light colored particles coupled strongly to the Higgs, namely the stops. Moreover, the Higgs sector in these models is extended to include an extra Higgs doublet, in the Minimal Supersymmetric extension of the SM (MSSM), and an additional extra singlet within the Next-to-Minimal Supersymmetric extension of the Standard Model (NMSSM). This implies, in general, that the Higgs boson couplings will depend strongly on the mixing of this particle with the additional neutral Higgs states and may present small deviations with respect to the SM ones. In particular, as emphasized above, the departures of the Higgs coupling to top quarks and the triple Higgs couplings from the SM values, may have an important impact on the di-Higgs production rate. This can also provide a probe to the nature of the electroweak phase transition due to its close connection to the triple Higgs coupling as pointed out in~\cite{Menon:2004wv,Grojean:2004xa,Noble:2007kk,Barger:2011vm,Chung:2012vg,Katz:2014bha,Curtin:2014jma,He:2015spf,Huang:2015tdv,Espinosa:2011ax,Profumo:2007wc,Profumo:2014opa,Kozaczuk:2015owa,Kotwal:2016tex,Morrissey:2012db,Chiang:2017nmu}.

The correlation of the di-Higgs production amplitude with the single Higgs production amplitude implies that the new physics contributions to the di-Higgs production channel are restricted by the Higgs production rate, which, as mentioned above, is bounded to be close to the SM prediction. This was stressed in Ref.~\cite{Batell:2015koa}, where they quantify the enhancements in the di-Higgs production due to stops while keeping gluon fusion single Higgs production rate, trilinear Higgs coupling ($\lambda_3$) and modification of the top quark Yukawa coupling with Higgs at or close to the SM values. Their results confirm that it is very difficult to achieve large deviations of the di-Higgs production at the LHC from the SM value under these constraints. In this work, we vary all three quantities within $2\sigma$ of the experimentally allowed ranges allowed by the
combined ATLAS-CMS analysis~\cite{Khachatryan:2016vau,Aad:2015zhl}, taking also into account the constraints coming from run-2 data.
We also extremize $X_t$ at each parameter space point to allow the maximal stop mixing parameters consistent with
theoretical and experimental constraints. We quantify the lightest stop mass dependence of the di-Higgs production cross section to show that fairly large deviations of the cross section are possible with these variations.

This work is organized as follows. In Sec. II, we develop a general understanding of heavy colored scalar particles contribution to the di-Higgs production cross section. As an example of  heavy colored particle, we take the stop contribution to the di-Higgs production process for the range of value of the couplings  allowed by the results of the combined ATLAS and CMS Higgs data at the  run-I of the LHC, and comment on possible modifications induced by run-II data. In Sec. III, we discuss the light stop one-loop contribution to $2h$ production. We show that when the gluon, as well as the bottom and top Higgs couplings are allowed to vary within the range consistent with the best fit values of $\kappa_i$,  the  $2h$ production may be greatly enhanced compared to the SM predictions. Moreover, we study the di-Higgs invariant mass distribution, showing that its study may lead to relevant information on the new particles contributing to the di-Higgs production rate. We reserve Sec. IV for our conclusions. In the Appendix~\ref{appendix1}, we present the form factors we use to perform the full one loop calculation.


%

 \section{Modifications of the gluon fusion Higgs and di-Higgs production cross sections}
 \label{sec:xsec}
 In this section, we consider the modification to double Higgs production cross section in the presence of light stops.
 We begin by writing down the stop mass matrix.
   \begin{align}\label{eqn:stopmassmatrix}
   \mathbf{M_{\tilde{t}}^{2}} =
   \left( \begin{array}{cc}
     m_{Q}^{2}+m_{t}^{2}+D_Q & m_{t}X_{t} \\
     m_{t}X_{t} & m_{U}^{2}+m_{t}^{2}+D_U\\
     \end{array} \right)
   \end{align}
 with $D_Q=m_z^2(T_3^t-Q^ts_W^2)\cos2\beta$, and $D_U = m_z^2Q^ts_W^2\cos2\beta $. The parameters $m_{Q}$ and $m_U$ are soft supersymmetry (SUSY) breaking mass terms of the left-handed and right-handed stops respectively, $X_t=A_t-\mu\cot\beta$ is the stop mixing parameter, $Q^t =2/3$ is the top quark charge, $T_3^t = 1/2$, $s_W$ is the sine of the weak mixing angle and $\tan\beta$ is the ratio of Higgs vacuum expectation values (VEVs). Neglecting the small contributions from $D$-terms ($m_z^2/3 \ll m_t^2, m_{Q,U}^2$) we obtain:
 \begin{align}\label{eqn:stopmasses}
  m_{\tilde{t}_{1,2}}^2&=\frac{1}{2}\big[m_{Q}^2+m_U^2+2m_t^2\mp \sqrt{(m_Q^2-m_U^2)^2+4m_t^2X_t^2}\big]
   \end{align}
  
\begin{figure}[ht!]
		\centering
		\includegraphics[width=1\textwidth]{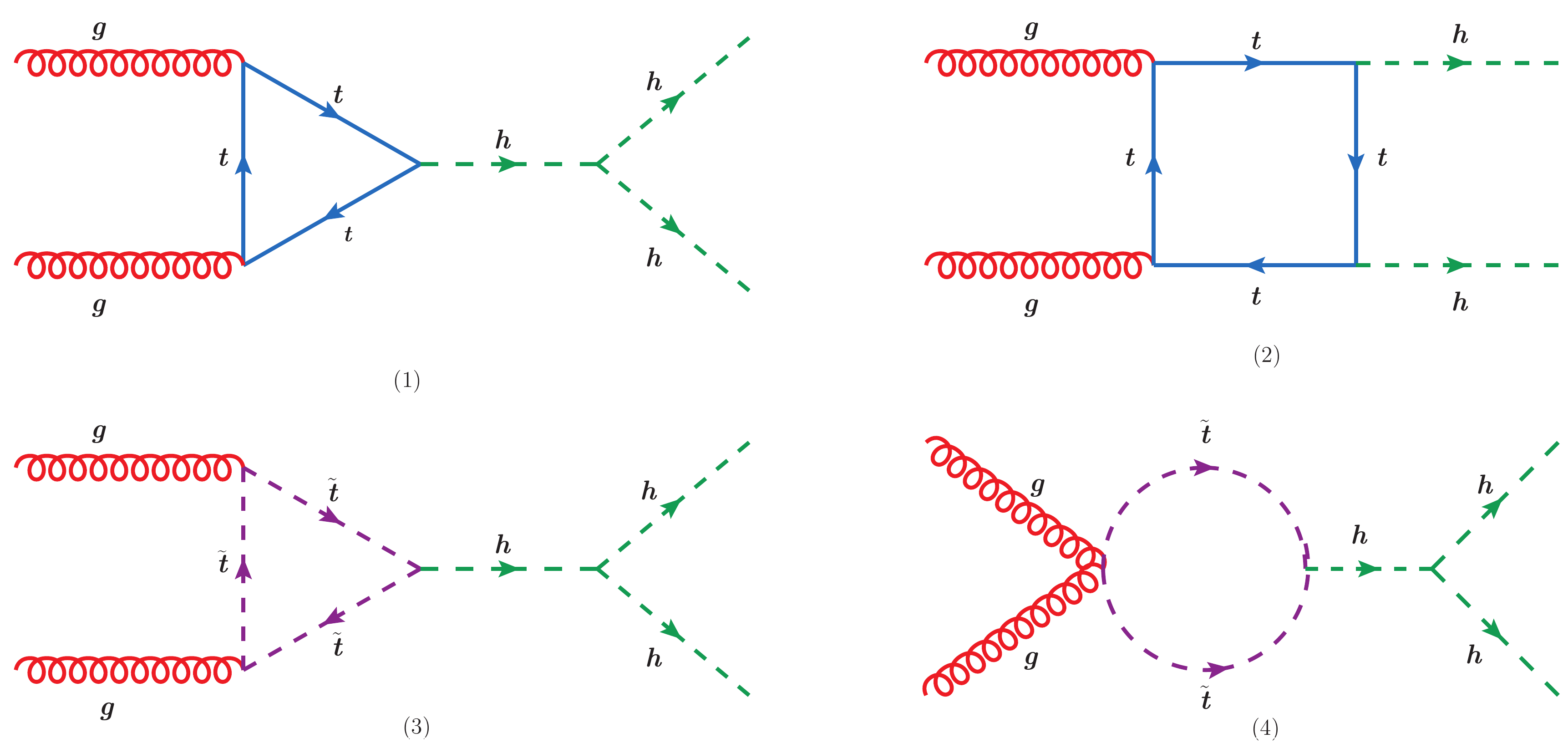}\\
		\includegraphics[width=1\textwidth]{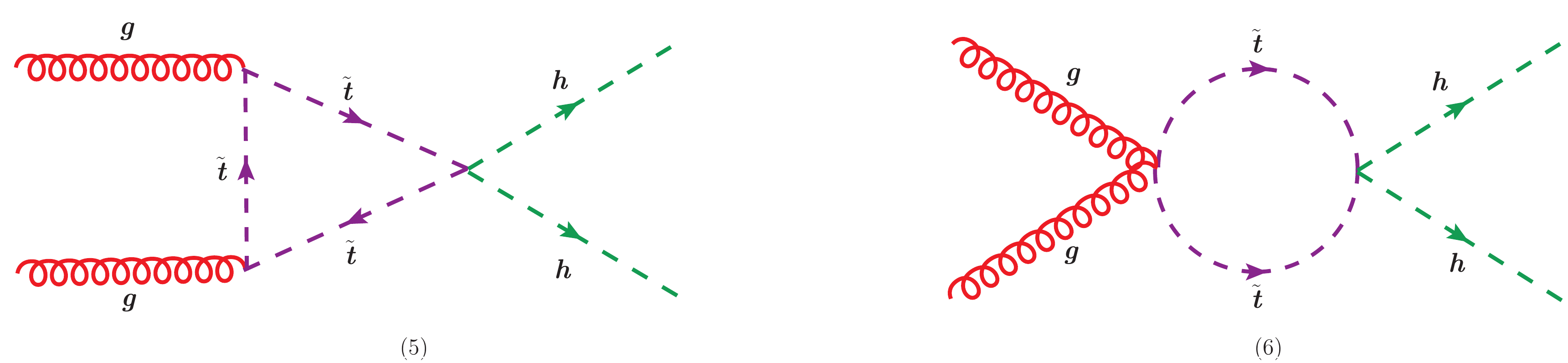}\\
		\includegraphics[width=1\textwidth]{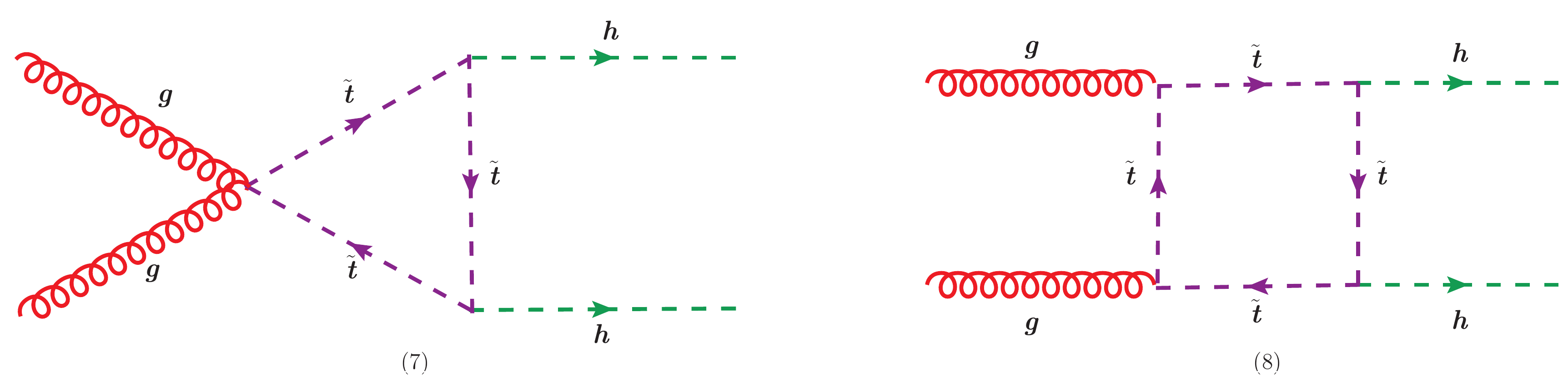}
		\caption{\label{fig:MSSM}MSSM 1-Loop diagrams of di-Higgs production.}
	\end{figure}

In the presence of light stops, in addition to the triangle and box diagrams with top-quarks in the loop, shown in Fig.~\ref{gghh_fulp}, there are new diagrams contributing to the double Higgs production at the leading order, shown in  Fig.~\ref{fig:MSSM}. 
Diagram (1) and (2) is the SM contribution, which may be modified by departures of 
the top-quark Yukawa coupling and the trilinear coupling with respect to the SM values. Diagrams (3) to (8) represent the stop contributions. While the dimensionful trilinear coupling of the Higgs to the stops has a strong dependence on the Higgs mixing parameter $X_t$, which can be larger than the stop masses, the  quartic coupling (bilinear in both the Higgs and stop fields) is fixed by the square of the top-quark Yukawa coupling. As the LHC has determined that the stops are significantly heavier than the top-quarks, diagrams (5) and (6) lead to only small contributions to the di-Higgs production rate, since there is no source of parametric enhancement associated with them. Diagrams (3), (4), (7) and (8) depend relevantly on the stop mixing parameter and tend to give the most relevant light stop contributions to the Higgs production rate. 

We have computed the leading order (LO) double Higgs production amplitudes depicted in Fig.~\ref{fig:MSSM}, finding agreement with the results presented in
Ref.~\cite{Belyaev:1999mx}. The Standard Model next to leading order (NLO) corrections have been computed in the literature in different approximations~\cite{Dawson:1998py,Frederix:2014hta,Maltoni:2014eza,Grigo:2014jma,Degrassi:2016vss,Grigo:2015dia,deFlorian:2016uhr,deFlorian:2015moa}. Recently, the full NLO QCD corrections
have become available~\cite{Heinrich:2017kxx}. Far less is known about the supersymmetric corrections. 
The NLO QCD corrections to the supersymmetric contributions have been calculated in the limit of vanishing external momenta\cite{Agostini:2016vze}, and they can be sizable with stops below the TeV scale. 

Due to the lack of a full NLO calculation in the supersymmetric case, in this work we will compute the cross section in MSSM at LO, then present its ratio to the SM LO value.
According to~\cite{Agostini:2016vze}, the K-factor in the MSSM can be larger than the K-factor in the SM, and therefore the ratio of the cross sections can be further enhanced by the NLO corrections.  In light of these results, our computation could be viewed as a conservative estimate of the possible enhancement due to supersymmetric particles. In order to numerically compute the supersymmetry effects on di-Higgs production we have modified the calculations presented in the public program MCFM-8.0~\cite{Campbell:2010ff}, by including the stop contributions and allowing also for possible modifications of the top-quark Yukawa coupling $y_t$ and the trilinear Higgs  coupling~$\lambda_3$. 

In order to generate relevant modifications to the double Higgs production cross section, the lightest stop mass should not be too far above from the weak scale. Therefore, the constraints on the stop masses coming from LHC searches put strong restrictions on the possible size of the supersymmetric contributions to the di-Higgs production process. Considering the direct production of stops at the LHC, the constraints on the stop masses depend strongly on the how stops decay, and on the masses of other particles in the decay chain. The most relevant constraints come from the region in which the difference between the stop and the lightest  neutralino masses are larger than the top-quark mass. In such a case, one would expect to find a significant decay of the stop into a top-quark and a neutralino. 

In the simplified models LHC considers, a stop decays one hundred percent either to a top and the lightest neutralino, or to a bottom and the lightest chargino, which then decay to a $W^+$ and the lightest neutralino. The final state is therefore a bottom, a $W^+$ and missing energy in both cases, but the kinematic distributions and efficiencies are different. The current constraints on the stop mass in this region of parameters are at least 500~GeV, and depend on the mass of the lightest neutralino,
becoming stronger for larger mass difference of the stop with the lightest neutralino~\cite{ATLAS:2017kyf,ATLAS:2017msx,Aaboud:2016tnv,Sirunyan:2017cwe,Sirunyan:2017kqq,CMS:2017vbf,Sirunyan:2017xse,CMS:2017qjo}. Moreover, the stop bound in the compressed regions is also of the order of 500~GeV~\cite{ATLAS:2017kyf,ATLAS:2017msx,ATLAS:2017tmd,CMS:2017odo}.

With more complicated decay chains, the constraints on stops could be weaker than the 500~ GeV limit  reported by the LHC. For example, in the presence of a light stau~\cite{Carena:2013iba,Carena:2012gp}, the decay chain of the stop is $\tilde{t} \, \rightarrow \, b \, \tilde{\chi}_1^{\pm}\, \rightarrow \, b \, \nu \, \tilde{\tau}\, \rightarrow b \, \nu \, \tau \tilde{\chi}_1^0$.  The final state is two b-jets, two $\tau$'s and missing energy. As $\tau$'s are difficult to detect at the LHC, the stop constraints in this scenario could be weakened significantly compared to that in the simple models described above. Another way to evade the constraints in the compressed region is to consider gauge mediation models~\cite{Giudice:1998bp}. In gauge mediation models, the lightest neutralino can decay to a photon and a gravitino. Then in the compressed region, all other decay products other than the photons are too soft, and the final states are two photons and missing energy. The current diphoton plus missing energy search is only focused on the high mass region of the squarks, and the limit for stops around 500 ~GeV or below is weak~\cite{ATLASCollaboration:2016wlb,Sirunyan:2017yse}. 

Long lived stops which dominantly decay through a RPV coupling $\lambda_{ijk}\bar{u}_i\bar{d}_j\bar{d}_k$ can have weaker constraints from the LHC as well~\cite{Csaki:2015uza}. In such scenario, the long lived stops can decay into a pair of down-type quarks, which lead to a displaced dijet final state. Then by recasting the 8~TeV data~\cite{Csaki:2015uza}, for $c\tau \sim$ 0.1 mm, stops lighter than 200 GeV can be allowed, and for $c\tau \sim$ 0.4 mm, stops around 400 GeV can be allowed. The heavy stable charge particle (HSCP) search in this scenario is weaker compared to the displaced dijet final state for low values of $c\tau$ ~\cite{Csaki:2015uza,CMS:2017bnk,Kopeliansky:2015gbi}.

Of course, the exact limit in the above three scenarios can be only obtained by doing a detailed recast of the current LHC data. The current recasting tools only include the data up to 2.3 $fb^{-1}$ of the 13 TeV Run~\cite{Drees:2013wra,Caron:2016hib}, and it is beyond the scope of this work to analyze the exact stop limit in those scenarios. In this analysis, we are going to consider stops as light as 300~GeV as a reference value, showing how the effects on the di-Higgs production cross section depend on the exact stop mass bound.  

In addition to direct constraints, light stops also modify the single Higgs production cross section in the gluon fusion channel, which is well measured at the LHC~\cite{Khachatryan:2016vau,Aad:2015zhl,ATLAS:2017bxr,CMS:2017rli}. The effective gluon Higgs coupling in the presence of light stops, and modified top Yukawa can be calculated from the low energy effective theory (EFT) approach. 
The leading contribution to the $gg\rightarrow h$ process can be obtained from the one-loop QCD beta functions
	 of the heavy particles~\cite{Ellis:1975ap,Shifman:1979eb,Kniehl:1995tn,Carena:2012xa,Essig:2017zwe}.  
	 The effects may be understood from the contribution of heavy particles to the gluon kinetic term, namely
	\begin{eqnarray}
	\mathcal{L}_{\mathrm{eff}}=-\frac{1}{4}G_{\mu\nu}^aG^{a\mu\nu}\sum_{i}\frac{\beta_{i}g_s^{2}}{16\pi^{2}}\log \frac{\Lambda^{2}}{m_{i}^{2}}
	\end{eqnarray}
	where $g_s$ is the strong coupling constant and $\beta_i$ is the contribution of the particle of mass $m_i$ to the QCD $\beta$ function. 
	If the masses of the loop particles depend on the Higgs VEV, 
	one can obtain the couplings of the SM-like Higgs by replacing the dependence on the VEV. by a dependence on the Higgs field, i.e. $m_{i} \rightarrow m_{i} (h)$.  If the particles are much heavier than the relevant energy scale $m_{h}$, then we can integrate out those particles and write the single Higgs production process, $gg\rightarrow h$ using a $1/m_i$ expansion of the effective Lagrangian. By substituting $h\rightarrow h+v$, the Taylor expansion of the  QCD effective Lagrangian on the Higgs field leads to 
	\begin{eqnarray}\label{EFT_Taylor_expansion}
	\mathcal{L_{\mathrm{eff}}}=-\frac{1}{4}G_{\mu\nu}^aG^{a\mu\nu }\sum_{i}\frac{\beta_{i}g_s^{2}}{16\pi^{2}}[\log \frac{\Lambda^{2}}{m_{i} (\upsilon)^{2}}&-&h\frac{\partial\log m_{i} (\upsilon)^{2}}{\partial\upsilon}\nonumber\\&-&\frac{h^{2}}{2!}\frac{\partial^{2}\log m_{i} (\upsilon)^{2}}{\partial\upsilon^{2}}+\cdots]
	\end{eqnarray}
	Therefore, one can get an approximation to the single and di-Higgs couplings to the gluon field strength. The effective Lagrangian for multi-Higgs couplings to gluons has been extended to $\text{N}^4\text{LO}$~\cite{Spira:2016zna}. However, these effective couplings are computed at zero momentum transfer and hence they lose validity, if the typical momenta are of the order or larger than the masses of the particles running in the loop. 
	
	 Let us stress that in single Higgs production the relevant scale is given by the Higgs mass and therefore this expansion leads to a good
	description of the effective coupling of the single Higgs to the gluon field strength for loop particle masses of the order of or above  the weak scale. In the case
	of double Higgs production, however, the relevant di-Higgs invariant mass scale may be much higher than the particle masses, and hence the effective field theory tends to fail for relatively light particles, leading to large corrections to the di-Higgs production process~\cite{Batell:2015koa,Dawson:2015oha}, as we discuss in the Appendix~\ref{EFTanalysis}. Hence, in our analysis, we shall use the full one-loop computation of the di-Higgs production rate.  
		
	From Eq.~(\ref{EFT_Taylor_expansion}), the contribution of the new particles to the linear coupling of Higgs to gluons may be obtained by the first order expansion of QCD effective Lagrangian can be written as:
	\begin{eqnarray}\label{2}
	\mathcal{L}_{gg\rightarrow h}=\frac{\alpha_{s}}{16\pi}G^{a}_{\mu\nu}G^{a\mu\nu}\frac{h}{\upsilon}\sum_{i}\beta_{i}\frac{\partial\log [\det{M_i^{\dagger}(\upsilon) M_i(\upsilon)}]}{\partial\log\upsilon}
	\end{eqnarray}
where we have grouped particles with the same quantum numbers in a singlet mass matrix $M_i$ and $\beta_{i}$ denotes their common
contribution to the QCD  beta function.\par

The contribution from stops to the single Higgs production rate has been previously considered in the literature. Including possible modifications in the Higgs coupling to tops, the modifications in $\kappa_g$ is given by~\cite{Djouadi:2005gj,Buckley:2012em,Carmi:2012in,Badziak:2016exn,Badziak:2016tzl}
  (see Appendix~\ref{EFTanalysis})
 \begin{eqnarray}\label{kappagmodifiedyt}
   \kappa_{g}= \kappa_t  + \frac{\kappa_t}{4} m_t^2 \left[\frac{1}{m_{\tilde{t}_1}^2}+ \frac{1}{m_{\tilde{t}_2}^2} - \frac{\tilde{X}_t^2}{m_{\tilde{t}_1}^2 m_{\tilde{t}_2}^2}\right].
      \end{eqnarray}
The value of $\kappa_t$ is governed, at tree-level, by the mixing between the $CP$-even Higgs bosons. In the MSSM, for instance, $\kappa_t \simeq \cos\alpha/\sin\beta$, $\tilde{X}_t^2 = X_t (A_t + \mu \sin\alpha/\cos\alpha)$, where $\alpha$ is the $CP$-even Higgs mixing angle. Close to the alignment (or decoupling) limit $\sin\alpha \simeq -\cos\beta$ and $\cos\alpha \simeq \sin\beta$, and for moderate values of $|\mu|$ and $\tan\beta$, $\tilde{X}_t^2$  is very well approximated by $X_t^2$.  Let us stress, however, that in the MSSM relevant deviations of $\kappa_t$ from one can only be obtained at low values of $\tan\beta$, for which the loop corrections are insufficient to bring the Higgs mass in agreement with observations for stops at the TeV  scale~\cite{Hahn:2013ria,Draper:2013oza,Bagnaschi:2014rsa,Vega:2015fna,Lee:2015uza,Bahl:2017aev}. In general, even for $\kappa_t = 1$, we will assume that the Higgs mass is not given by the MSSM  relations, but fixed by additional D-terms, that  could arise in gauge extensions of the MSSM~\cite{Batra:2003nj},  or F-terms, like happens in the NMSSM~\cite{Ellwanger:2009dp}. In the    NMSSM, similar relations between $\tilde{X}_t^2$ and $X_t^2$ are obtained, and small deviation of $\kappa_t$ from one may be obtained for both heavy as well as light scalar singlets~\cite{Badziak:2016exn,Badziak:2016tzl}.
      
In addition, let us comment that $\kappa_t = 1.1$ in general implies the presence of additional, relatively light, non-standard Higgs bosons. The additional $CP$-even Higgs could lead to a resonant double Higgs production if their masses are larger than 250~GeV (for an analysis of the resonant production in singlet extensions of the SM, see Ref.~\cite{Dawson:2015haa,No:2013wsa}). For instance, in the MSSM (with additional D-terms to fix the Higgs mass and induce a significant Higgs mixing), for a 350 GeV heavy $CP$-even Higgs, with low $\tan\beta \simeq 1$, and relatively heavy stops, the gluon fusion cross section is of the order of several pb~\cite{Bahl:2016brp}. The branching ratio for such heavy Higgs to a pair of SM Higgs bosons, in the absence of light charginos or neutralinos, is around a few tens of percent, which leads to a cross section for $p p \rightarrow H \rightarrow h h$ higher than the nonresonant double Higgs production. In this case, we expect to see a resonance in the $m_{hh}$ distribution. 

When the heavy Higgs mass is increased to about 500~GeV, the gluon fusion cross section is around a pb, and the branching ratio of $H\rightarrow h h $ decreases to about a few percent as the $t\bar{t}$ channel opens up. Then the resonant production rate is comparable to the nonresonant production rate. Because of the destructive interference between the resonant double Higgs production diagram and the box diagram, we expect a dip-peak structure in the $m_{hh}$ distribution at the parton level. Whether this structure is visible at the LHC depends on how large the cross section is, which depends on model parameters as $\tan\beta$, and $m_{\tilde{t}}$, and the detector resolution.  

We should also stress that these results are strongly model dependent. In the NMSSM, when light singlets are present, or when there are large splittings between the $CP$-even and $CP$-odd Higgs bosons, the dominant Higgs decay of the heavy $CP$-even Higgs boson is into non-standard Higgs states, and the resonant production of pairs of SM-like Higgs bosons is highly suppressed~\cite{Carena:2015moc,Kling:2016opi,Badziak:2016tzl}. Moreover, even with the MSSM Higgs sector, in the presence of light neutralinos or charginos, the decay branching ratio into pair of SM-like Higgs bosons  could be highly suppressed~\cite{Badziak:2016exn}. In this article, we shall concentrate on the nonresonant production of SM-like Higgs bosons, and analyze the impact of the mixing with additional Higgs bosons via the modifications of the top-quark Yukawa and trilinear Higgs self couplings.
      
As can be seen from Eq.~(\ref{kappagmodifiedyt}), a small enhancement in the Higgs coupling to tops, as currently  allowed by  data~\cite{Khachatryan:2016vau,Aad:2015zhl}, would not only enhance the top-quark Yukawa coupling, but also the stop contribution to the gluon fusion cross section.  Let us stress however, that the run~I indications of a high value of $\kappa_t$~\cite{Khachatryan:2016vau,Aad:2015zhl} have not been confirmed by the current run II data~\cite{ATLAS:2017ovn,ATLAS:2016ldo,CMS:2017vru} and hence in the following we shall consider only small variations of this coupling.  Moreover, the stop effects may be significantly enhanced for large values of $X_t$, which could also lead to 
a reduction of the Higgs coupling to gluons $\kappa_g$, within the range allowed by the run~I best fit values. However, a very large $X_t$ might also affect the Higgs vacuum stability~\cite{Frere:1983ag,Claudson:1983et,Falk:1995cq,Kusenko:1996jn,Chowdhury:2013dka,Blinov:2013fta}. In this paper, following the results of Ref.~\cite{Blinov:2013fta}, we shall use the approximate bound 
    \begin{eqnarray}\label{HgsVcmStb}
      A_t^2\le  \left(3.4+0.5\frac{\lvert 1-r\lvert}{1+r}\right)(m_{Q}^2 + m_{U}^2)+60\left( \frac{m_z^2}{2} \cos(2\beta) + m_A^2 \cos^2\beta \right)
      \end{eqnarray}
      when $\mu$ is small, $A_t\simeq X_t$.  Here $r = m_{Q}^2/m_{U}^2$ and
      the last term represents the impact of the $CP$-odd Higgs mass. In general, we will take a conservative approach and neglect the second term, containing the dependence
      on $m_A$ and $m_z$,  and consider the above bound on $A_t$ as a bound on $X_t$, as would approximately arise from Eq.~(\ref{HgsVcmStb}) at large values of $\tan\beta$ and moderate values of $\mu$ and $m_A$.  However, we will
      also discuss the impact of considering the above bound on $A_t$ for low values of $\tan\beta$ and moderate values of $m_A$ and $\mu$.

In summary, we have calculated the modifications to double Higgs production with a modified version of MCFM in the presence of a light stop, modified top Yukawa and Higgs trilinear couplings. The mass of the light stop, the stop mixing angle, and the modifications to the couplings are subject to direct stop searches, precision Higgs signal strength measurements, and the vacuum stability. The numerical results for the double Higgs production cross section and the current experimental constraints will be discussed in the next section.

\section{Collider Phenomenology}

\begin{figure}[tbp]
\centering
\includegraphics[width = .8\textwidth]{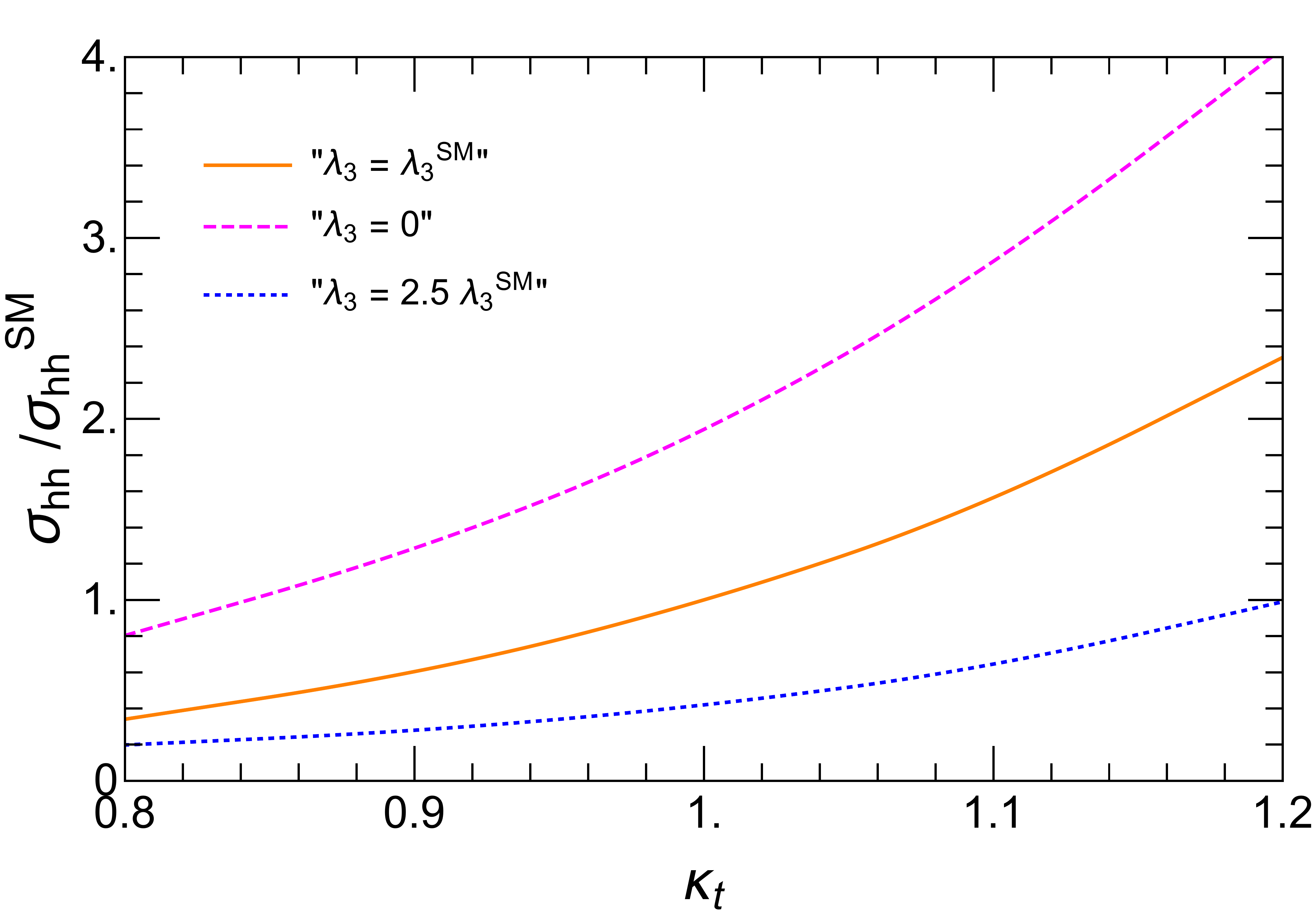}
\caption{\label{fig:xsecvskt}Di-Higgs production cross section in the absence of stops, as a function of the top-quark Yukawa coupling, $\kappa_t$, for different values of the Higgs trilinear coupling $\lambda_3$. Here, we have $\kappa_t=\kappa_g$.}
\end{figure}

The collider phenomenology depends strongly on the precise stop masses, the stop mixing angle and the values of the top-quark Yukawa and trilinear Higgs couplings. In Fig.~\ref{fig:xsecvskt}, we compute the variation of the di-Higgs production cross section in the absence of stops. As it is clear from this figure, even a mild variation of the top quark Yukawa coupling, $\kappa_t = 1.1$, can lead to an increases of the cross section by 50~percent.
The reason for this is that the contribution to the SM amplitude associated with the box diagram, which increases quadratically with $\kappa_t$, is about a factor~2.5 larger than the one associated with the triangle diagram at the 2~$m_{t}$ threshold, 
which increases only linearly with $\kappa_t$, and interferes 
destructively with the box amplitude. 

We also show the variation of the cross section with a modification of the trilinear Higgs coupling. For $\kappa_t =1$, the value of $\lambda_3 = 2.5~\lambda_3^{\rm SM}$ maximizes the destructive interference between the box and triangle diagram amplitudes, and hence leads to a general reduction of the di-Higgs production cross section. On the contrary, for small values of $\lambda_3 \simeq 0$, only the box diagram contributes, and hence the cross section is not only enhanced with respect to the SM case, but depends quartically on the top quark coupling $\kappa_t$. Di-Higgs production cross section values of the order of 4~times the SM value may be obtained for the maximal variations of $\kappa_t$ and
$\lambda_3$ considered in Fig.~\ref{fig:xsecvskt}.

\begin{figure}[tbp]
\centering
\includegraphics[width = .50\textwidth]{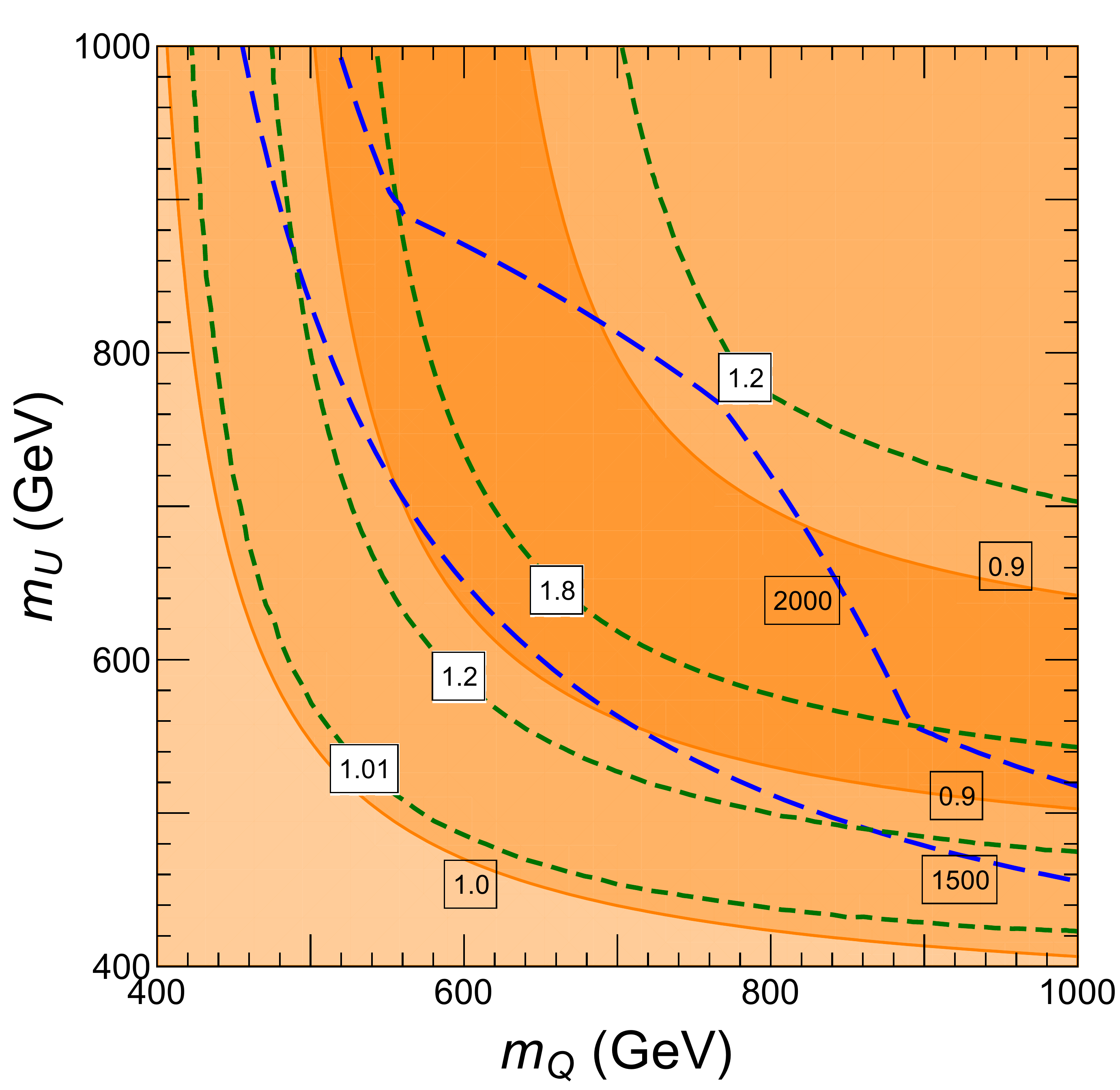}\hfill
\includegraphics[width = .50\textwidth]{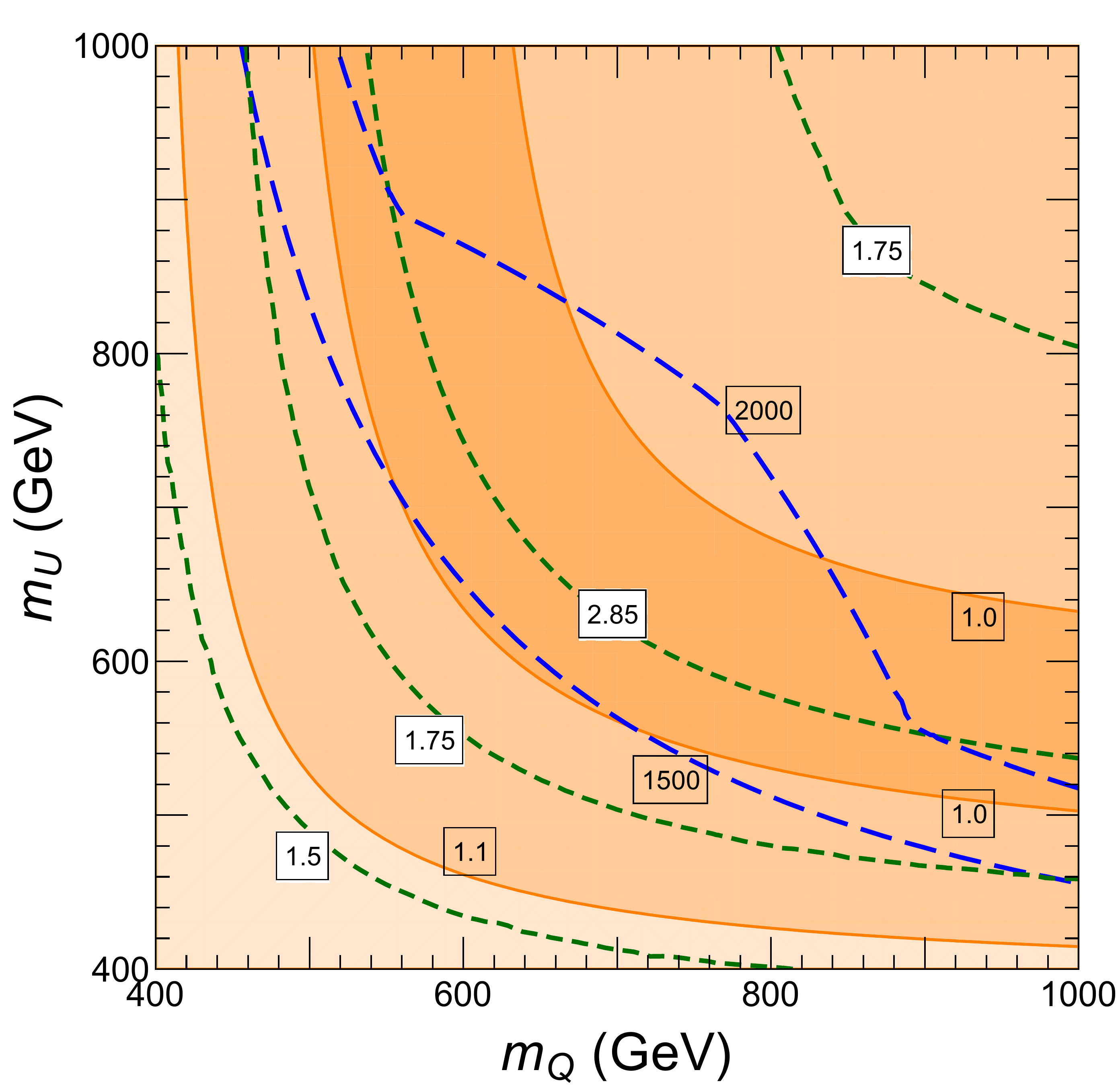}\\
\includegraphics[width = .50\textwidth]{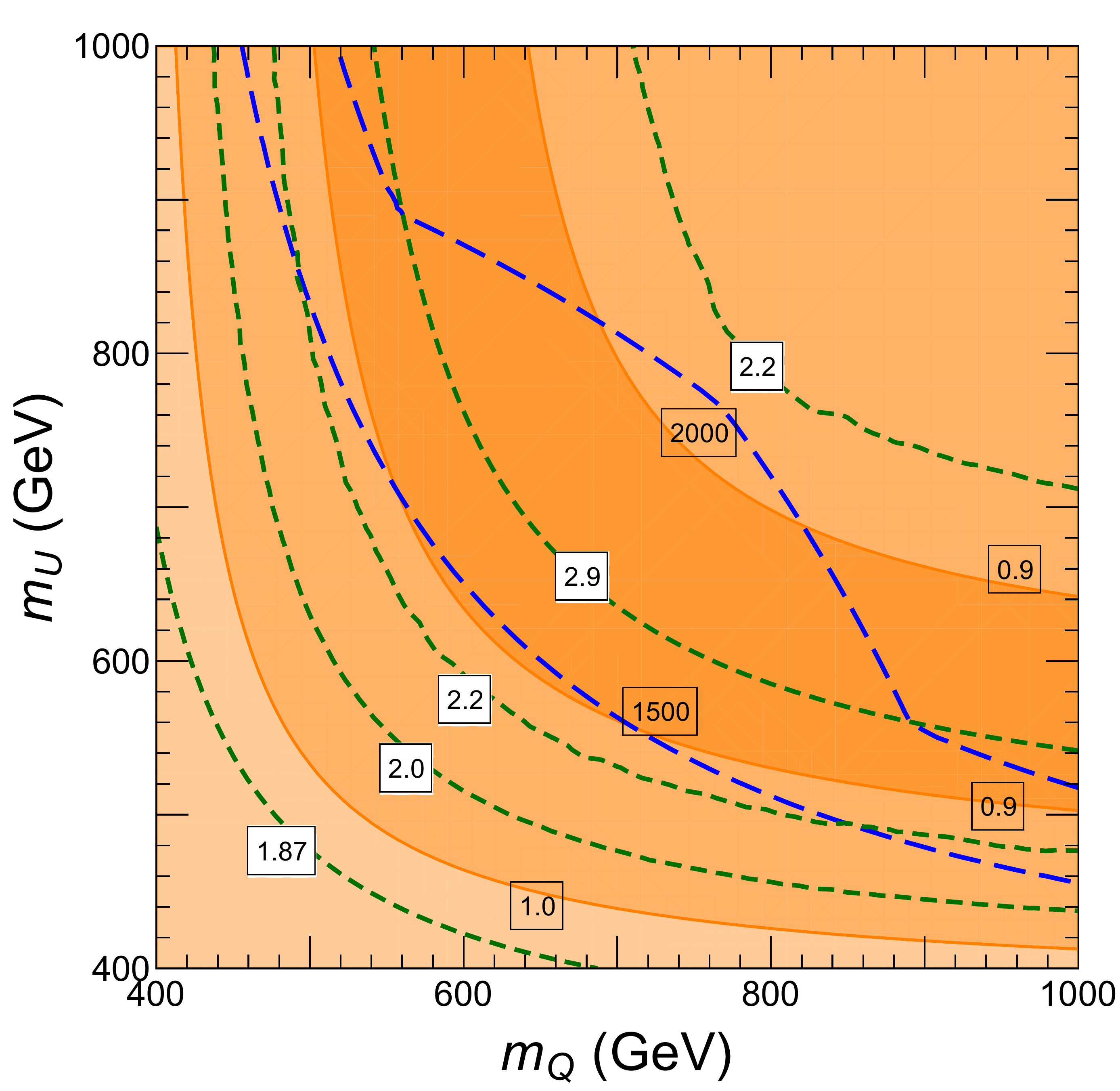}\hfill
\includegraphics[width = .50\textwidth]{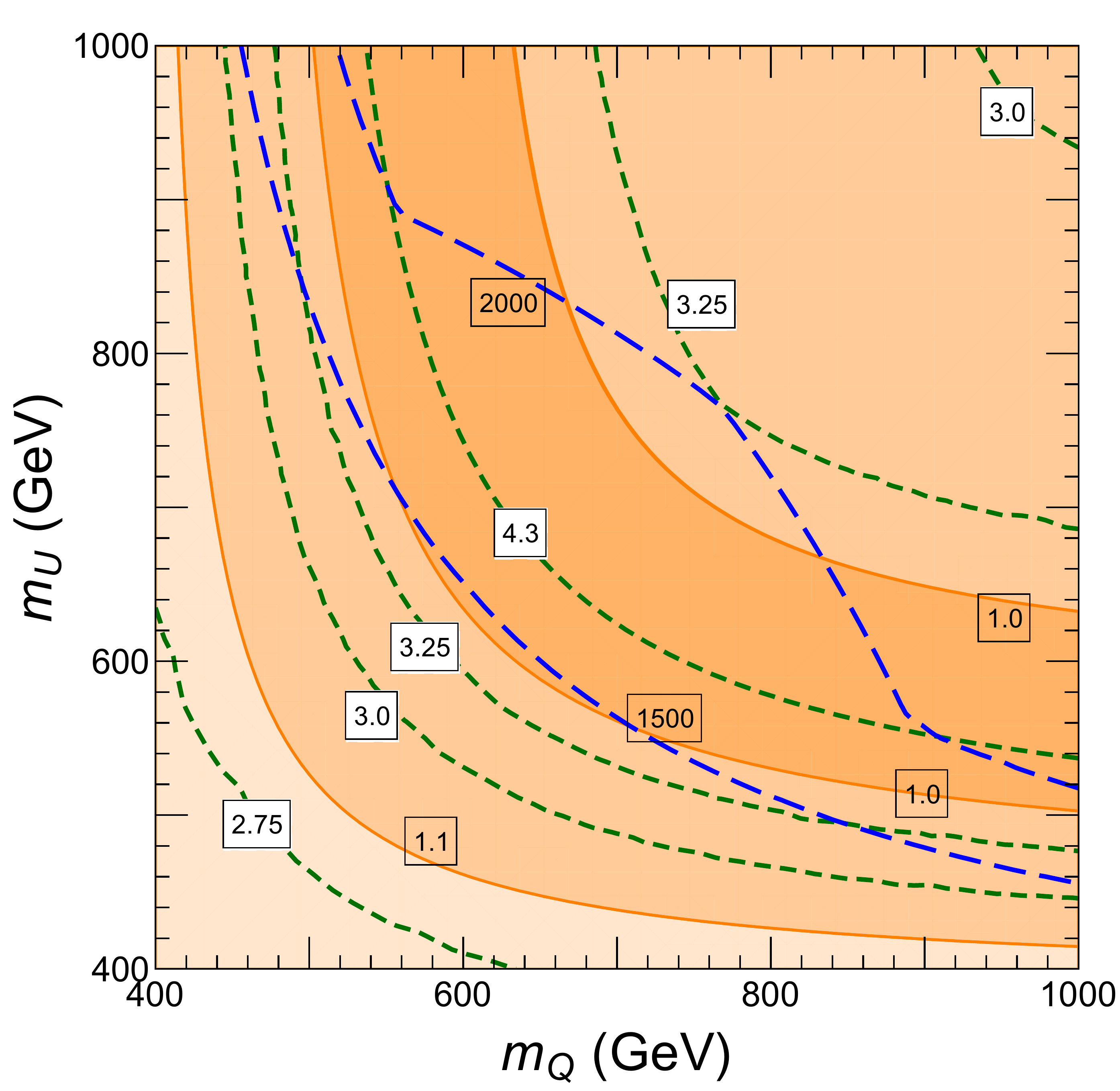}
\caption{\label{fig:xsec400} Double Higgs production cross section, considering a stop mass lower bound of 400~GeV. In the left panels, $\kappa_t$ = 1, while in the right panels $\kappa_t$ = 1.1. The first and the second row correspond to values of $\delta_3=0$ and $-1$ respectively. The green dashed contours show the total di-Higgs production cross section normalized to the SM value. 
The darkest (second darkest, lightest) orange regions represent values of $0.8 \ (0.9, 1.0) <\kappa_g < 0.9 \ (1.0,1.1)$ in the left panel, as shown on their boundaries. In the right panel, the darkest (second darkest, lightest) orange regions represent values of $0.9~(1.0, 1.1)~<\kappa_g~< 1.0~(1.1,1.2)$.
The blue contours show the maximum value of $|X_t|$ that is consistent with the stop mass bound and  vacuum stability constraints at each point.}
\end{figure}

\begin{figure}[tbp]
	
	\centering
	\includegraphics[width = .49\textwidth]{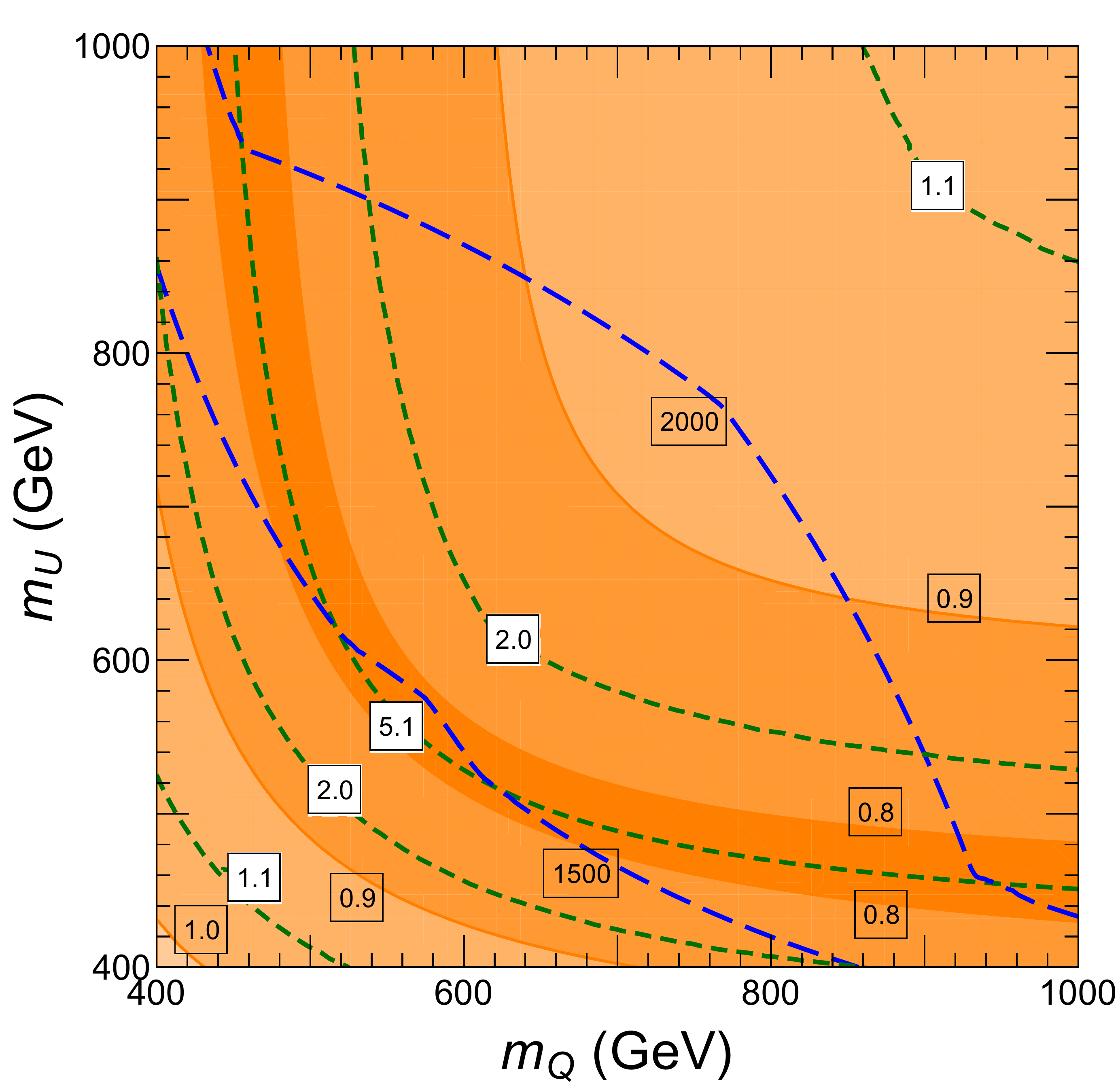}
	\hfill 
	\includegraphics[width = .49\textwidth]{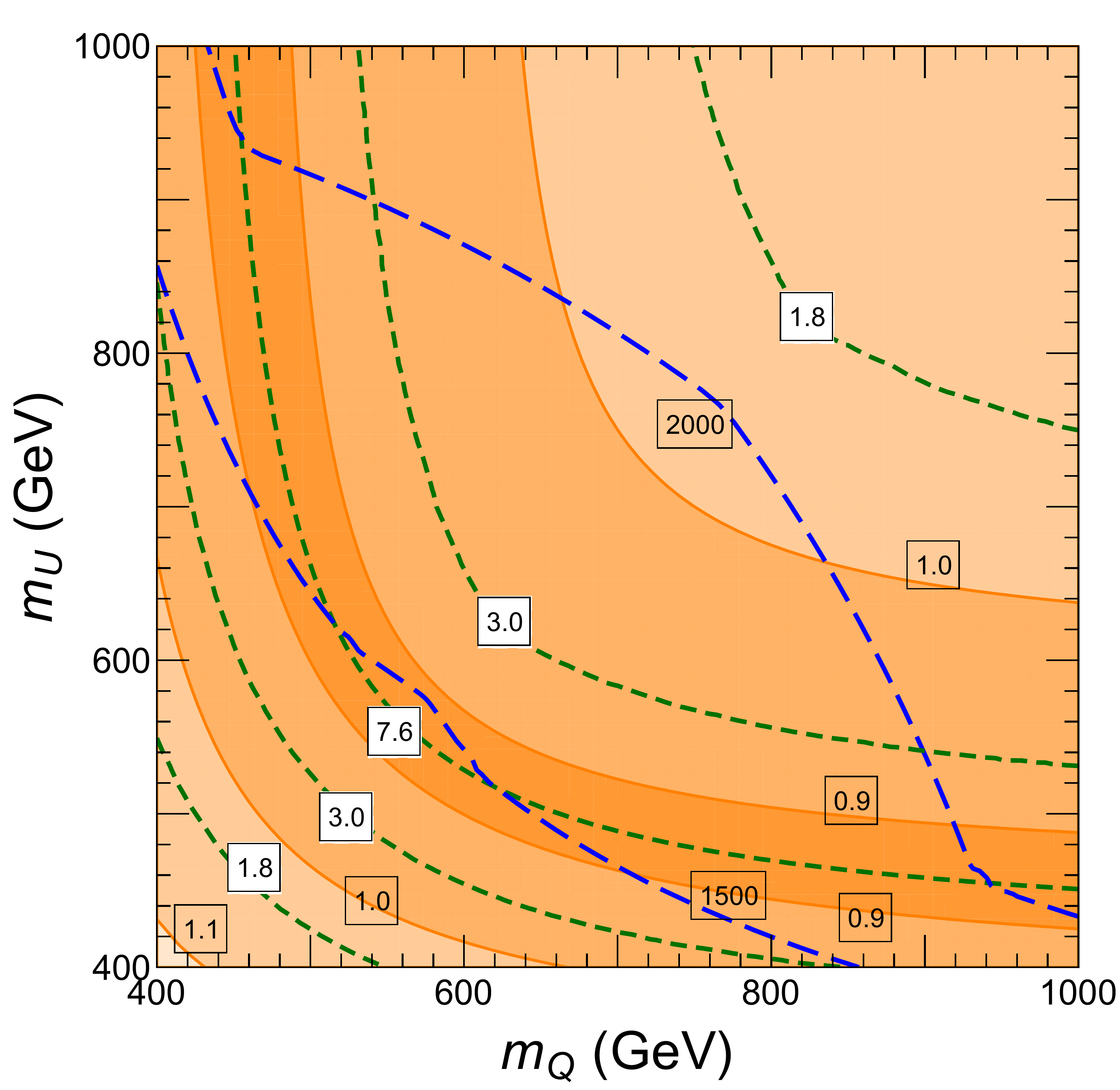} \\
	\includegraphics[width = .49\textwidth]{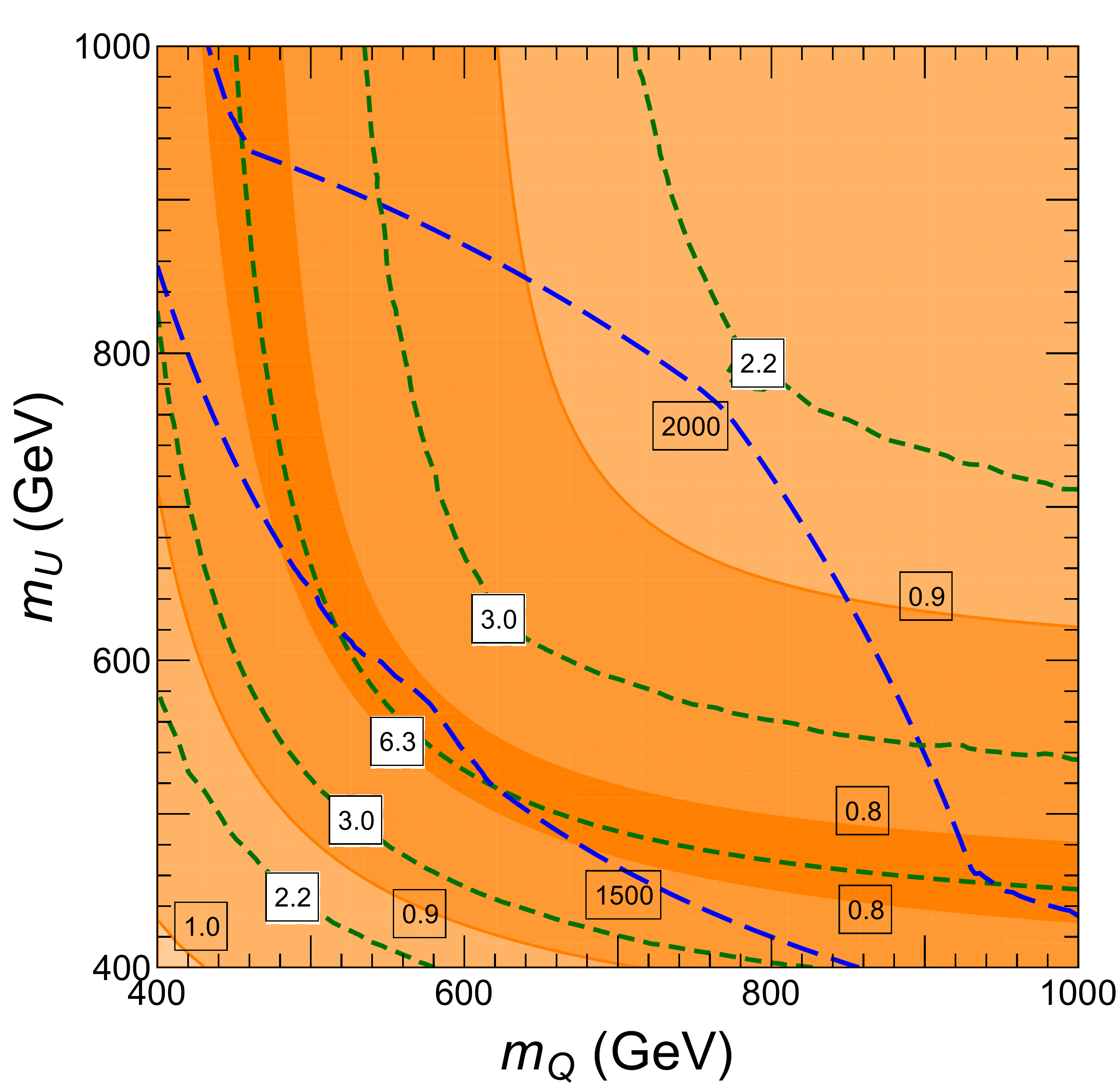}
	\hfill 
	\includegraphics[width = .49\textwidth]{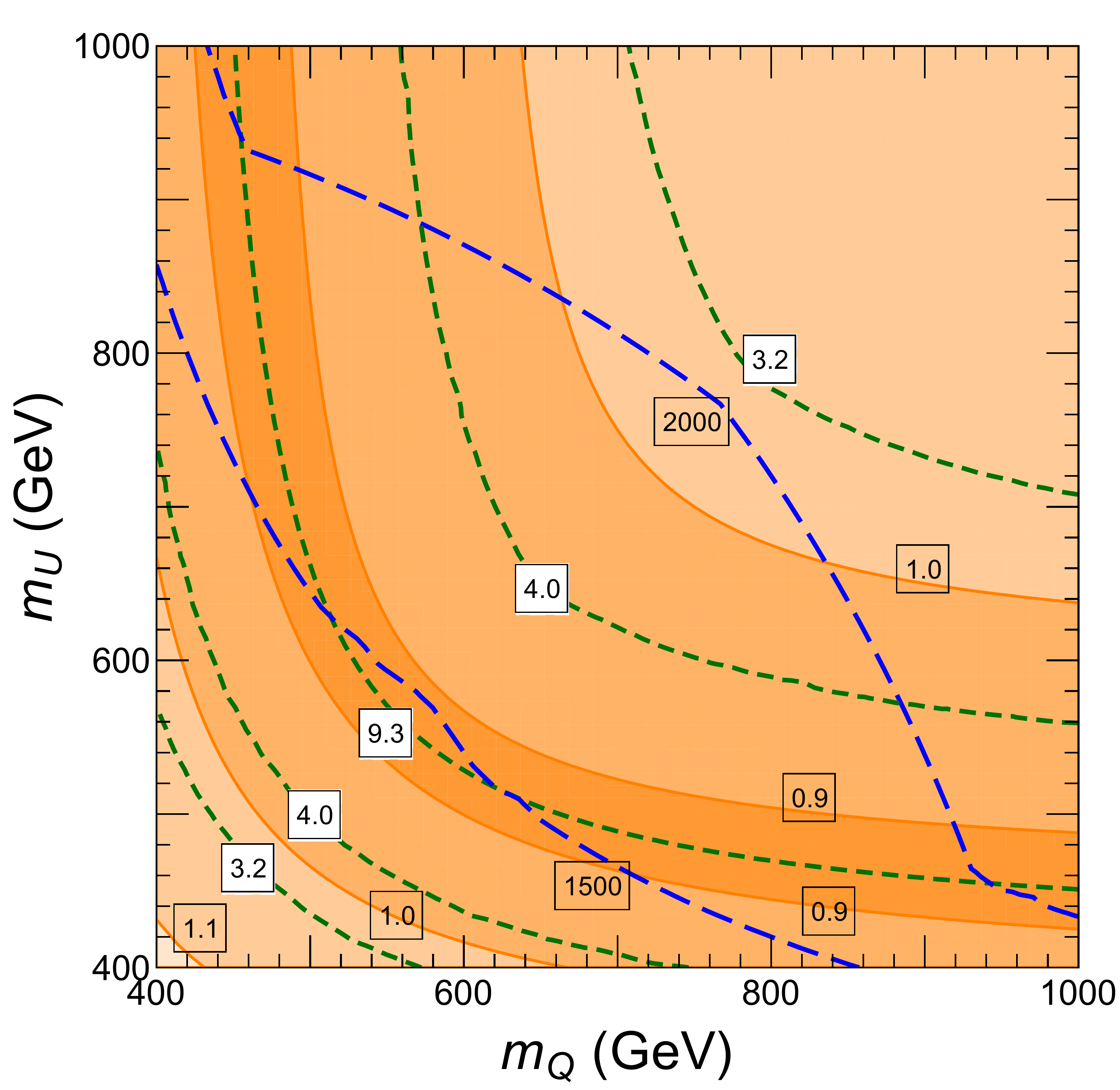}
	\caption{\label{fig:xsec300} Same as Fig.~\ref{fig:xsec400}, but considering a stop mass lower bound of 300~GeV. The darkest (second darkest, lightest) orange regions represent values of $0.7~(0.8, 0.9) <\kappa_g < 0.8~(0.9,1.0)$ in the left panel, as shown on their boundaries. In the right panel, the darkest (second darkest, lightest) orange regions represent values of $0.8~(9.0, 1.0)~<\kappa_g~< 0.9~(1.0,1.1)$.}
\end{figure}

\begin{figure}[tbp]
	\centering
	\includegraphics[width = .49\textwidth]{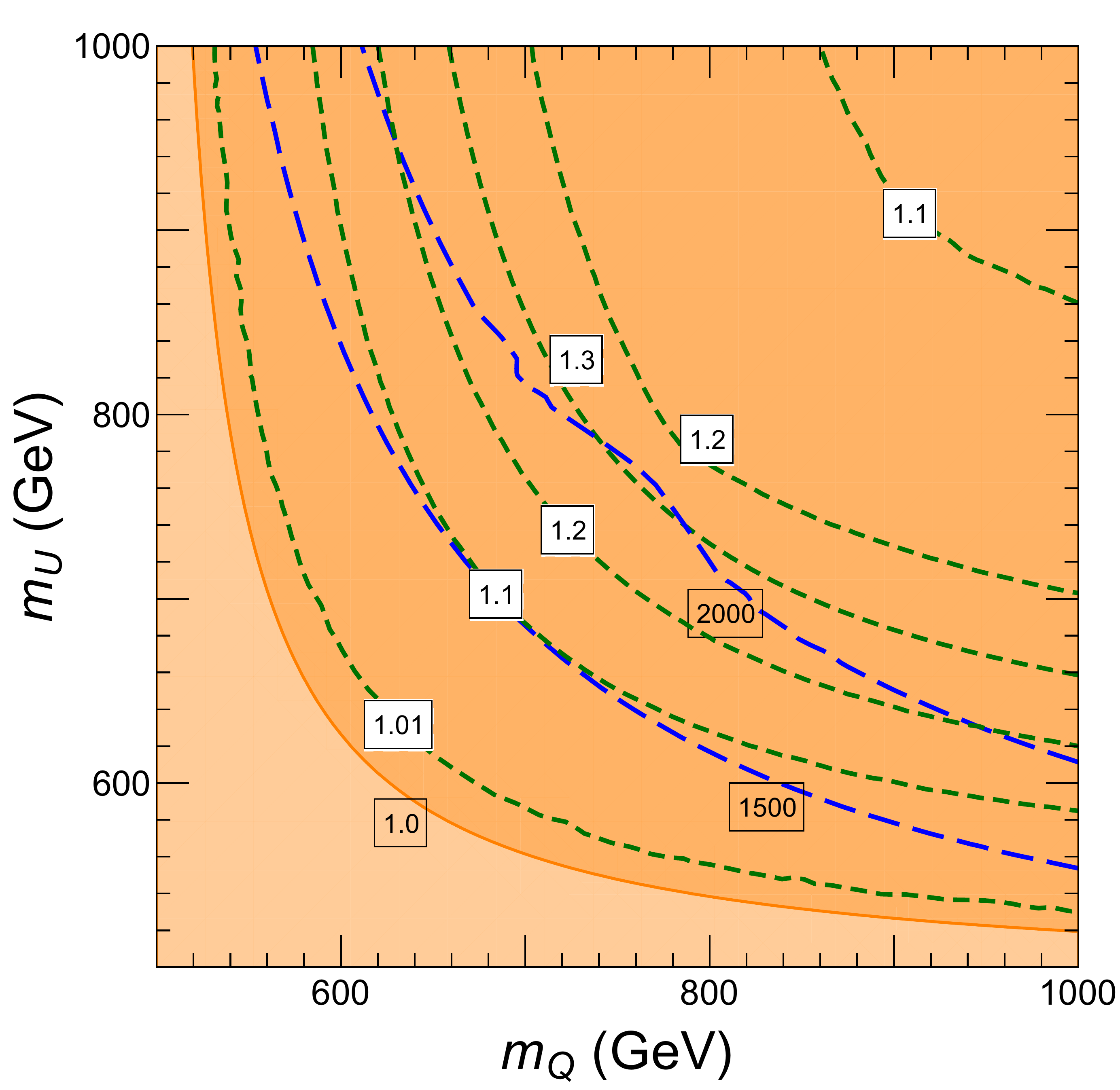}
	\hfill 
	\includegraphics[width = .49\textwidth]{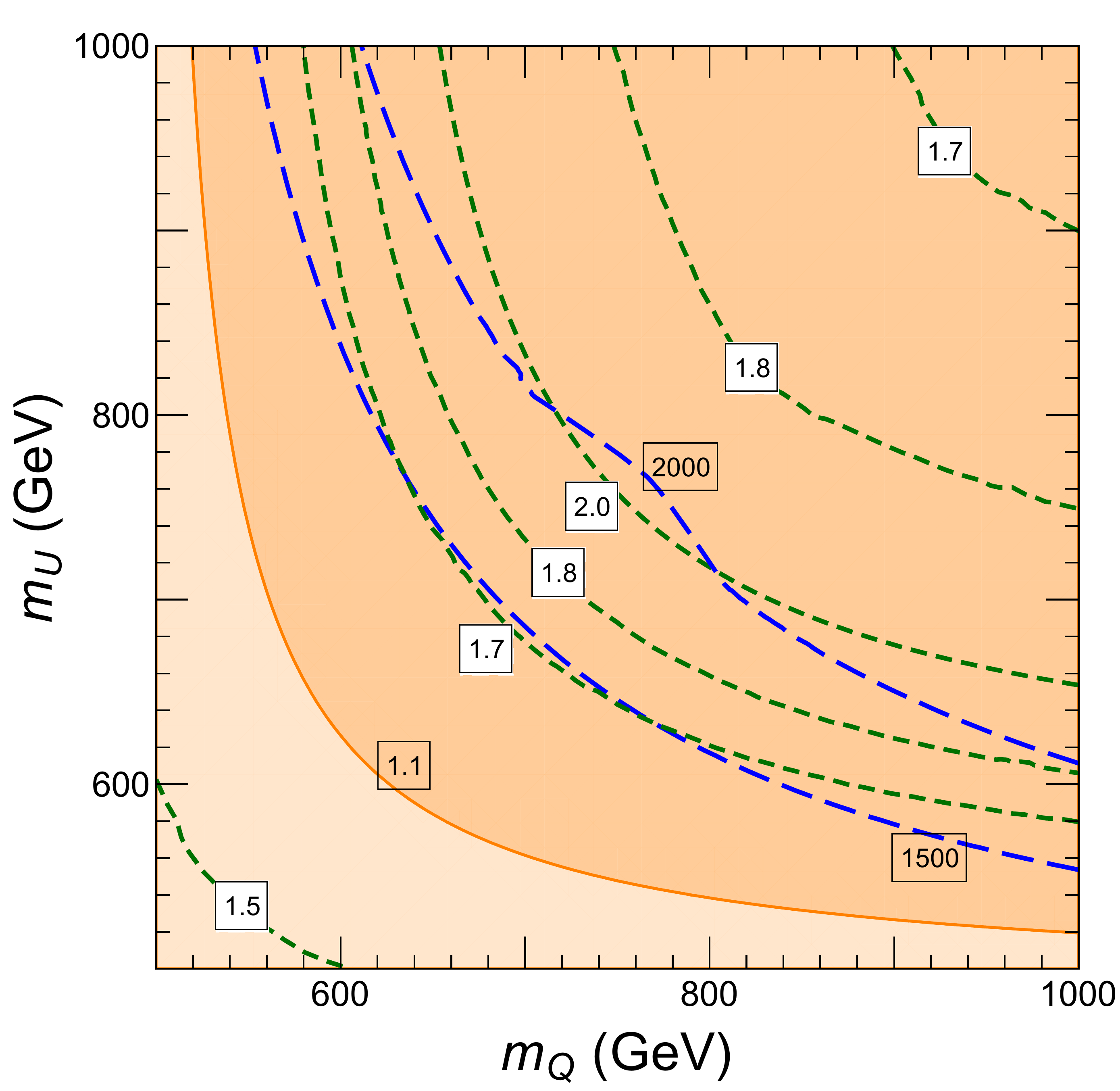} \\
	\includegraphics[width = .49\textwidth]{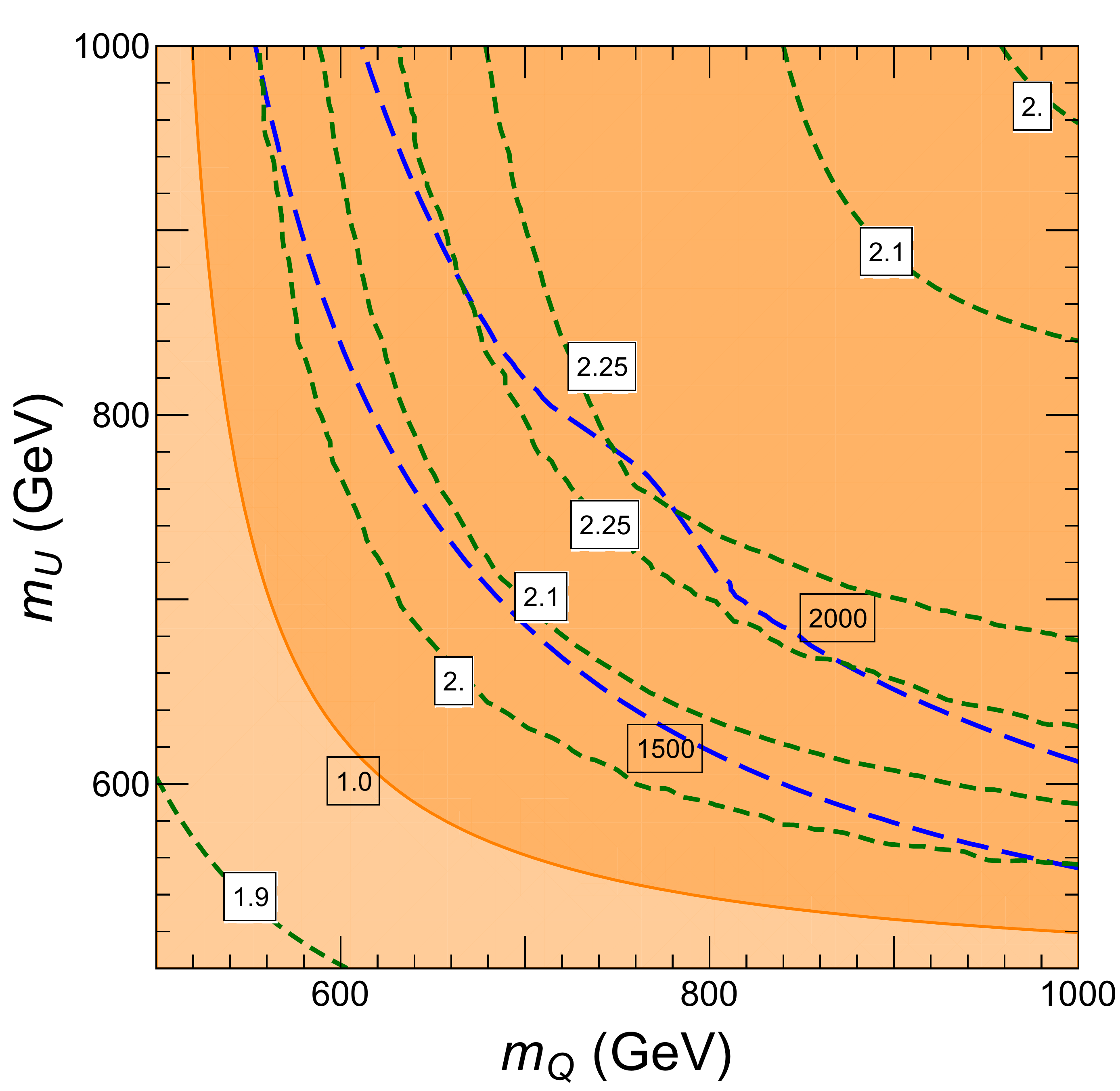}
	\hfill 
	\includegraphics[width = .49\textwidth]{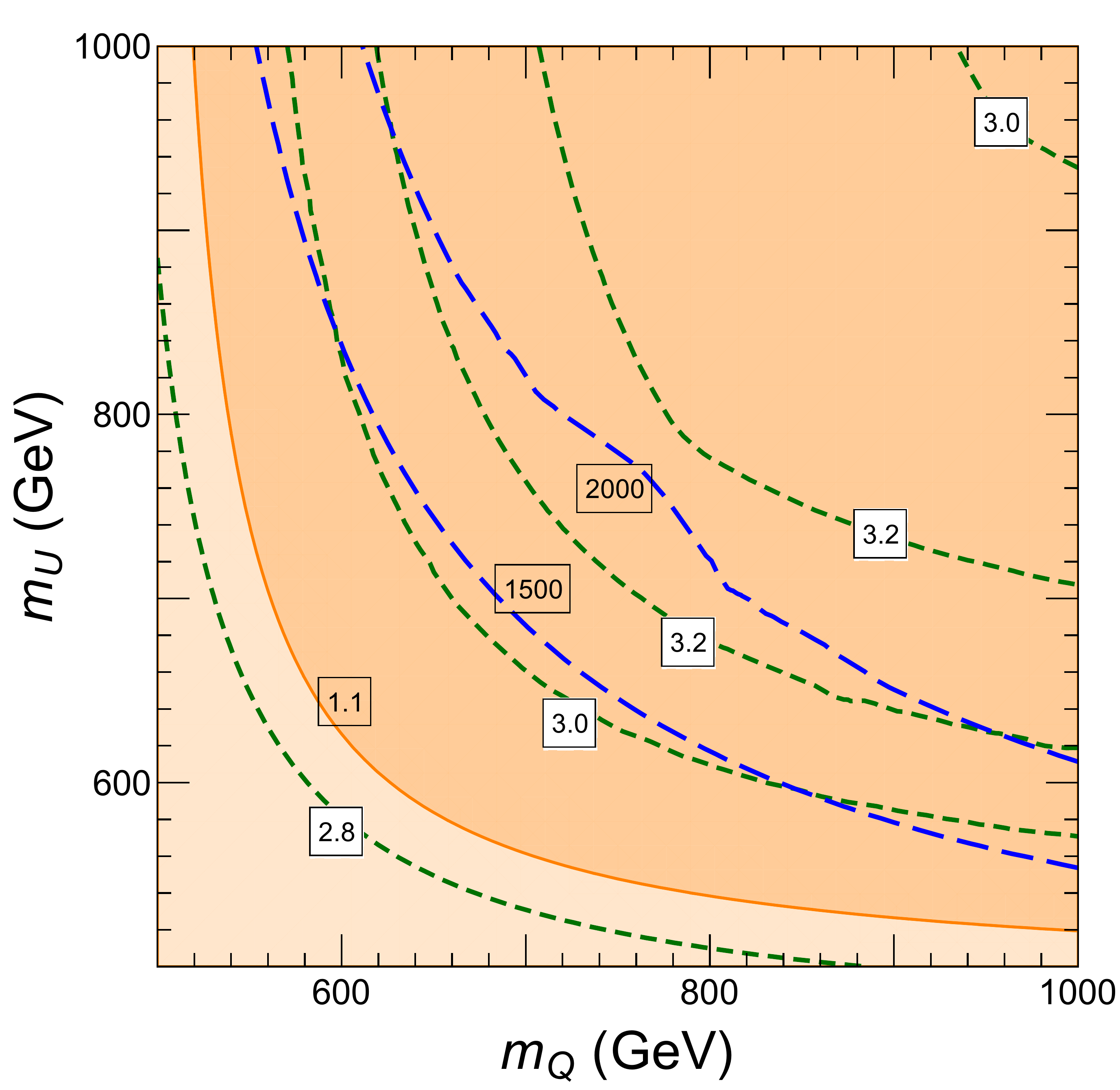}
	\caption{\label{fig:xsec500}Same as Fig.~\ref{fig:xsec400}, but considering a stop mass lower bound of 500~GeV. The darkest (lightest) orange regions represent values of $0.9~(1.0) <\kappa_g < 1.0~(1.1)$ in the left panel, as shown on their boundaries. In the right panel, the darkest (lightest) orange regions represent values of $1.0~(1.1)~<\kappa_g~< 1.1~(1.2)$.}
\end{figure}

\begin{figure}[tbp]
	\centering
	\includegraphics[width = .50\textwidth]{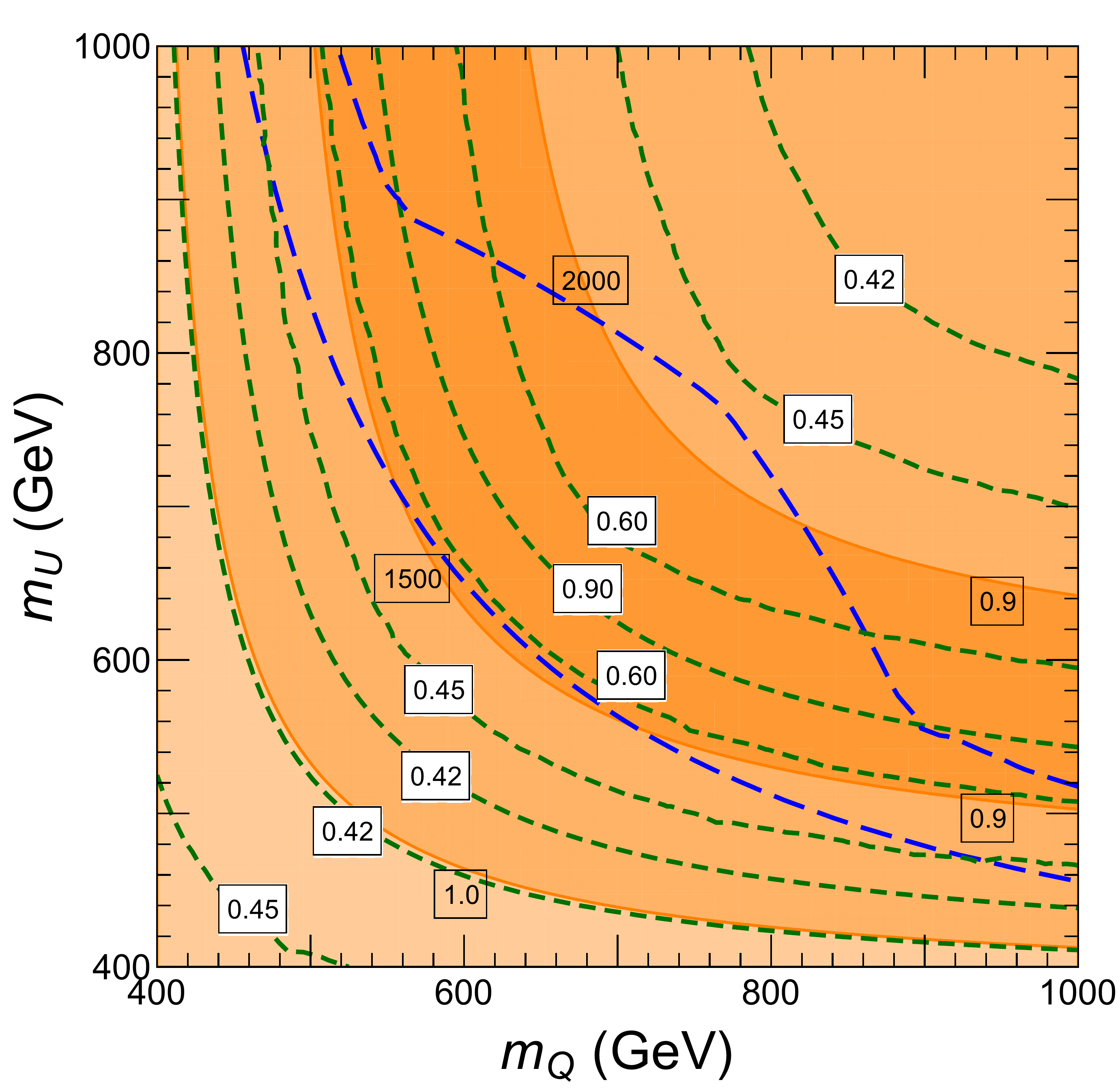}\hfill
	\includegraphics[width = .50\textwidth]{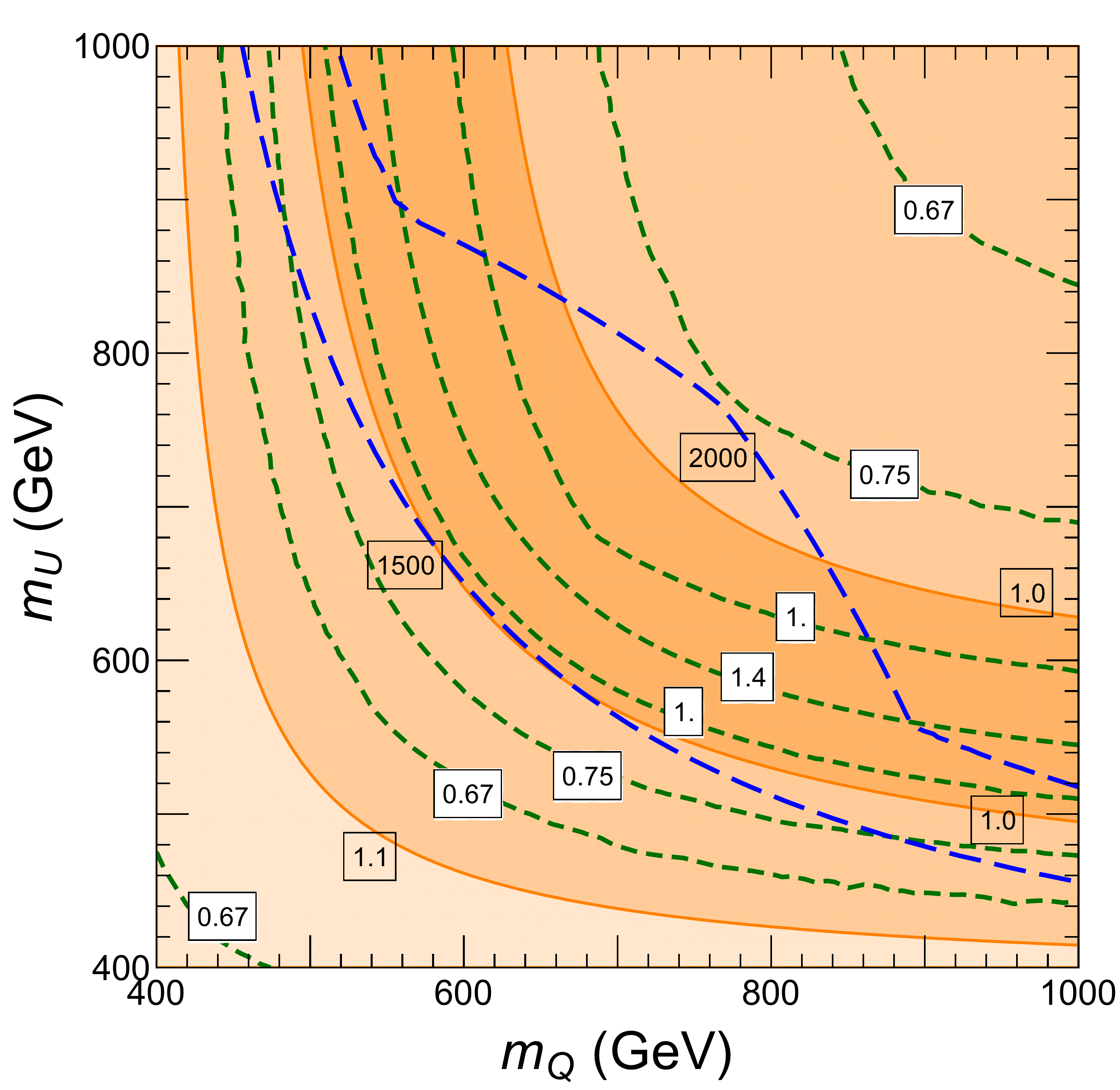}
	\caption{\label{fig:xsec40015} Same as Fig.~\ref{fig:xsec400}, but for $\delta_3=1.5$. Left panel corresponds to $\kappa_t=1$ and right panel corresponds to $\kappa_t=1.1$.}
\end{figure}

In the Figs.~\ref{fig:xsec400},~\ref{fig:xsec300},~\ref{fig:xsec500} and~\ref{fig:xsec40015}, we show the results for the double Higgs cross section in the presence of light stops. For each values of $m_Q$ and $m_U$, we calculated the largest value of $|X_t|$ that can be allowed by a lower bound on stop mass and a stable Higgs vacuum, with  a Higgs vacuum expectation value of $v = 246$~GeV. The lower bound on the stop masses used in Figs.~\ref{fig:xsec400},~\ref{fig:xsec300},~\ref{fig:xsec500} and~\ref{fig:xsec40015} are $400$~GeV, $300$~GeV,$500$~GeV, and $400$~GeV respectively. Then we use the previously mentioned modified version of MCFM-8.0 to calculate the double Higgs production cross section, which is normalized to the SM value, as shown by the green dashed contours. For the stability condition, we decided to be conservative and ignore the $m_A$ and $m_z$ dependence in Eq.~(\ref{HgsVcmStb}). The dependence on $m_A$ of the vacuum stability bound on $X_t$, and of the resulting double Higgs production cross section, will be discussed later.

We also calculate the single Higgs production cross section in the gluon fusion channel, as shown in the orange regions. The left panels in all three figures correspond to a value of the top-quark Yukawa coupling normalized to the SM value, $\kappa_t=1.0$, while the right panel corresponds to $\kappa_t=1.1$. The modification of the triple Higgs coupling is defined as $\delta_3=(\lambda_3 - \lambda_3^{\rm SM})/\lambda_3^{\rm SM}$. The first and last row in each of the Figs.~\ref{fig:xsec400},~\ref{fig:xsec300} and~\ref{fig:xsec500} corresponds to $\delta_3 = 0$ and $\delta_3=-1$. The latter is used as an example to demonstrate the effect on the di-Higgs production in the case of no destructive interference between the triangle and the box diagrams. Such values of $\delta_3$ can be realized in a scalar singlet extension of the SM Higgs sector, like the NMSSM, as demonstrated in~\cite{Huang:2015tdv}. Finally, Fig.~\ref{fig:xsec40015} shows the case $\delta_3 = 1.5$, in which, as stressed above, there is large destructive interference between the triangle and box diagram amplitudes. 

Assuming the Higgs top-quark and  triple Higgs couplings acquire SM values, the double Higgs cross section can be as large as 1.8 times the SM production cross section. When a small enhancement of the Higgs coupling to top-quarks is allowed, for instance $\kappa_t = 1.1$, the double Higgs production cross section can be as large as about~3 times the SM cross section with a 400 GeV light stop. Allowing the modifications of the Higgs trilinear coupling, for $\delta_3 = -1$ and $\kappa_t = 1$, similar enhancement of the di-Higgs production rate is obtained. When both modifications are considered together, i.e., $\kappa_t = 1.1$ and $\delta_3 = -1$, then the values of more than~4 times the SM di-Higgs production rate may be obtained as shown in the bottom right panel of Fig.~\ref{fig:xsec400}.

An important property that may be also extracted from Fig.~\ref{fig:xsec400}, is the strong correlation between the modification of the di-Higgs production cross section and the one of the gluon fusion single Higgs production rate. A large modification in the di-Higgs production indicates a large value of the stop mixing parameter $X_t$, which decreases the gluon fusion single Higgs production rate, Eq.~(\ref{kappagmodifiedyt}). Values of the Higgs coupling to gluons somewhat lower than 0.9 times the SM value are required to obtain the largest corrections to the di-Higgs production rate for $\kappa_t = 1$, while for $\kappa_t = 1.1$,  the required values are $0.9 \leq \kappa_g \leq 1$.  These values
of $\kappa_g$ are consistent with the combined fit to the run-I Higgs data, that leads to a preference for lower values of $\kappa_g \simeq 0.81^{+ 0.13}_{-0.11}$~\cite{Aad:2015zhl}. 

The situation changes drastically for different values of the lightest stop mass. For instance, as shown in Fig~\ref{fig:xsec300}, if stop masses of order 300 GeV would be allowed, di-Higgs production cross section of about 5 times the SM cross section could be obtained, even assuming no variation of the top-quark Yukawa and trilinear Higgs couplings, and values of order 6 to 8 times the SM production rate if either $\kappa_t = 1.1$ or $\delta_3 = -1$. Values of order 9 to 10~times
the SM di-Higgs production rate may be obtained for $\kappa_t = 1.1$ and $\delta_3 = -1$. On the other hand, as shown in Fig.~\ref{fig:xsec500}, if the lightest stop mass were 500~GeV or above, for $\kappa_t=1$ and $\delta_3= 0$, only small modifications of the order of 30~percent, with respect to the SM case would be allowed. Larger values of the cross section may be obtained for top-quark Yukawa and trilinear Higgs couplings that differ from the SM values. The enhancements of more than 3 times the SM rate may be obtained for $\kappa_t = 1.1$ and $\delta_3 = -1$.

The previously discussed correlation of the single and di-Higgs production rate remains approximately true for the different values of the stop masses. For stop masses of order of 300~GeV (500~GeV), the maximal modifications of the di-Higgs production cross section  demand corrections to the Higgs
coupling to gluons of somewhat less than 0.8 (1.0) times the SM value for $\kappa_t = 1$ and 0.9 (1.1) times the SM value for $\kappa_t = 1.1$. Similar values of $\kappa_g$, $\kappa_t$ and $\delta_3$ tend to be associated with similar corrections to the di-Higgs production cross section for
the large values of the stop mixing considered in Figs.~\ref{fig:xsec400},\ref{fig:xsec300},\ref{fig:xsec500} and~\ref{fig:xsec40015}.

Figure~\ref{fig:xsec40015} corresponds to $\delta_3=1.5$. As emphasized before, in the SM case, this value corresponds to a maximal reduction, due to the destructive interference between the triangle and the box diagrams, of the di-Higgs production~\cite{Frederix:2014hta,deFlorian:2015moa}. This value can also be realized in singlet scalar extensions of the SM Higgs sector, like the one present in the NMSSM, and is strongly correlated with the obtention of strongly first order phase transition,  which is of particular interest from the perspective of EW baryogenesis, as shown in~\cite{Noble:2007kk,Barger:2011vm,Katz:2014bha,Huang:2015tdv,Chen:2017qcz,Lewis:2017dme}. In the absence of stops and $\kappa_t = 1$, the double Higgs cross section falls to~$\sim 0.4$ times that of the SM value~\cite{Frederix:2014hta,deFlorian:2015moa}, making statistical significance too low even at the end of the LHC run. In  Fig.~\ref{fig:xsec40015}, the left panel shows that the addition of light stops, with the lighter one having a mass of $\sim 400$ GeV, can more than double this value to make it about $0.9$ times the SM cross section. In the right panel, we show that in addition to a $\sim 400$ GeV stop, small modifications ($\sim 10\%$) of $\kappa_t$ can increase the double Higgs cross section to about $40\%$ above the SM value.

Finally, it is worth emphasizing that the comparison of Figs.~\ref{fig:xsecvskt},\ref{fig:xsec400},\ref{fig:xsec300} and~\ref{fig:xsec500} show that,  due to the dominance of the box diagram, in general, for positive corrections of the di-Higgs cross section,   the modification of the top-quark and trilinear Higgs couplings have an approximate multiplicative effect in the modification of the cross section induced by light stops.  The cross sections obtained for light stops and modified couplings are approximately given by the ones obtained for light stops and SM values of the couplings times the ones obtained by modifying the couplings in the SM case, shown in  Fig.~\ref{fig:xsecvskt}. Also, for Figs.~\ref{fig:xsecvskt},\ref{fig:xsec400},\ref{fig:xsec300},~\ref{fig:xsec500} and~\ref{fig:xsec40015}, as we go to the region of high values for both $m_Q$ and $m_U$ (of about or more than 1 TeV), the vacuum stability constraints on  $X_t$ constrain the lightest stop mass to values larger than the minimum allowed one, and therefore the di-Higgs production cross section results become independent of the allowed lightest stop mass. Thus, the mass of the lighter stop increases for higher $m_Q$ and $m_U$, leading to the decoupling of the stop effects.

We also compare the full one-loop calculation(solid lines) with the EFT calculation (dashed lines) in Fig.~\ref{fig:EFTcomparison} (for the detailed EFT calculation, see Appendix~\ref{EFTanalysis}). We chose $\kappa_t = 1$ for orange, red and green lines, and $\kappa_t = 1.1$ for the blue lines. The value of $\kappa_g$ increases monotonously along each line except the orange lines with increasing mass of the lightest stop. The range of $\kappa_g$ values are from $0.85$ to $0.98$ for the red, $0.77$ to $0.98$ for the green and $0.90$ to $1.08$ for the blue line. For a given lightest stop mass and $\kappa_t$, the $\kappa_g$ value is same for EFT and one-loop case. The orange plots have $\kappa_g=1$ by definition. All lines are plotted taking $m_Q=m_U$ and $\delta_3=0$. 

\begin{figure}[tbp]
	\centering
	\includegraphics[width = 0.7\textwidth]{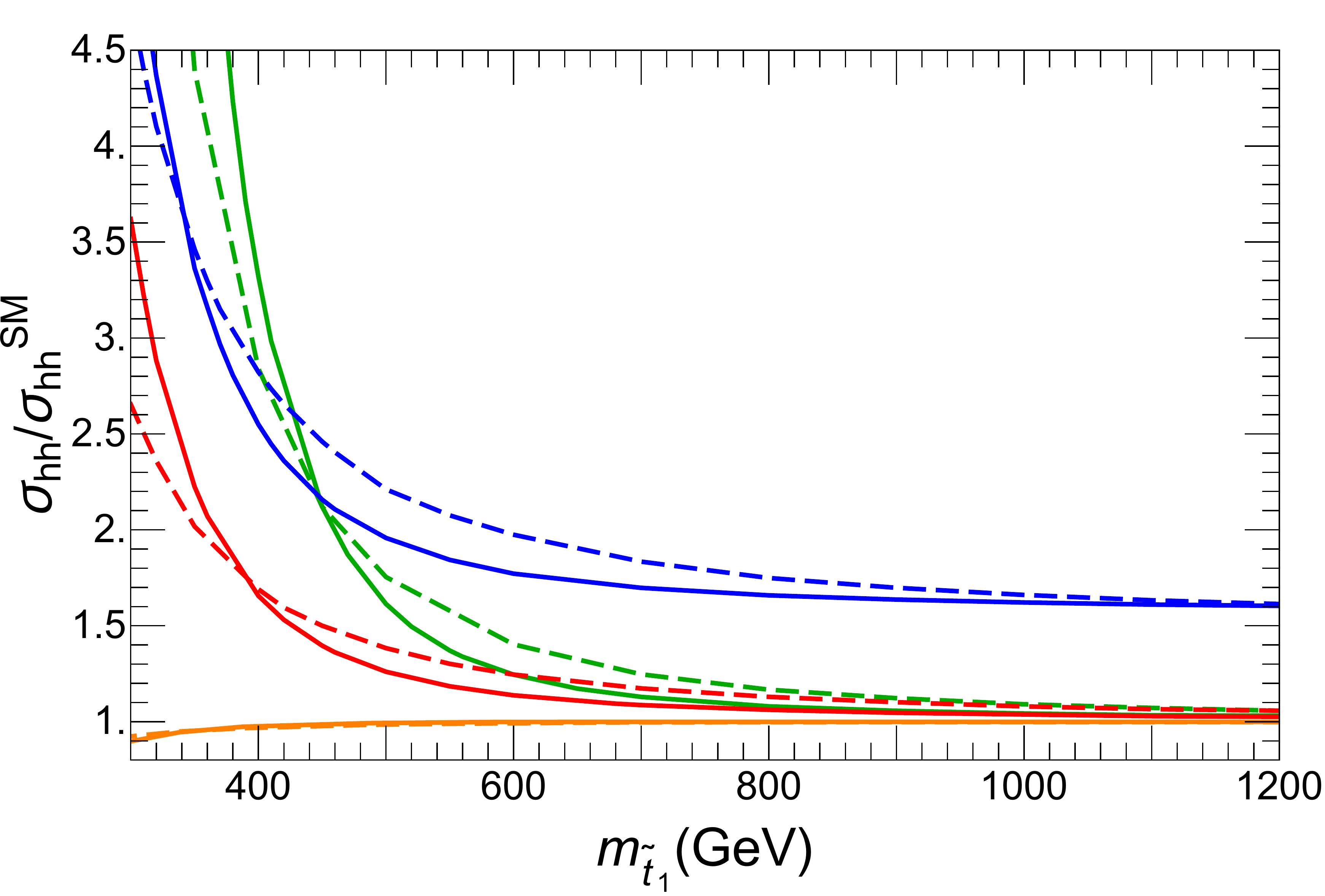}
	\caption{\label{fig:EFTcomparison}Di-Higgs production cross section normalized to the SM value using the full one loop calculation (solid lines) and the EFT calculation (dashed lines) as a function of the lightest stop mass for $m_Q=m_U$ and $\delta_3=0$. $\kappa_t$ is chosen to be 1 for the orange, red and green lines, and 1.1 for the blue lines. For the red and the blue lines, $X_t^2$ is chosen to saturate the vacuum stability condition as in Eq.~(\ref{HgsVcmStb}), neglecting the $m_A$ and $m_z$ terms. For the green lines, $X_t^2$ is chosen to saturate the vacuum stability condition with $m_A = 350$~GeV, $\mu = 400$~GeV, and $\tan\beta$ = 1. For the orange line, $X_t^2$ is chosen to be $m_{\tilde{t}_1}^2 + m_{\tilde{t}_2}^2$ to keep $\kappa_g = 1$. The range of $\kappa_g$ values for each of the lines are given in the text.}
\end{figure}

\begin{figure}[tbp]
	\centering
	\includegraphics[width = 0.7\textwidth]{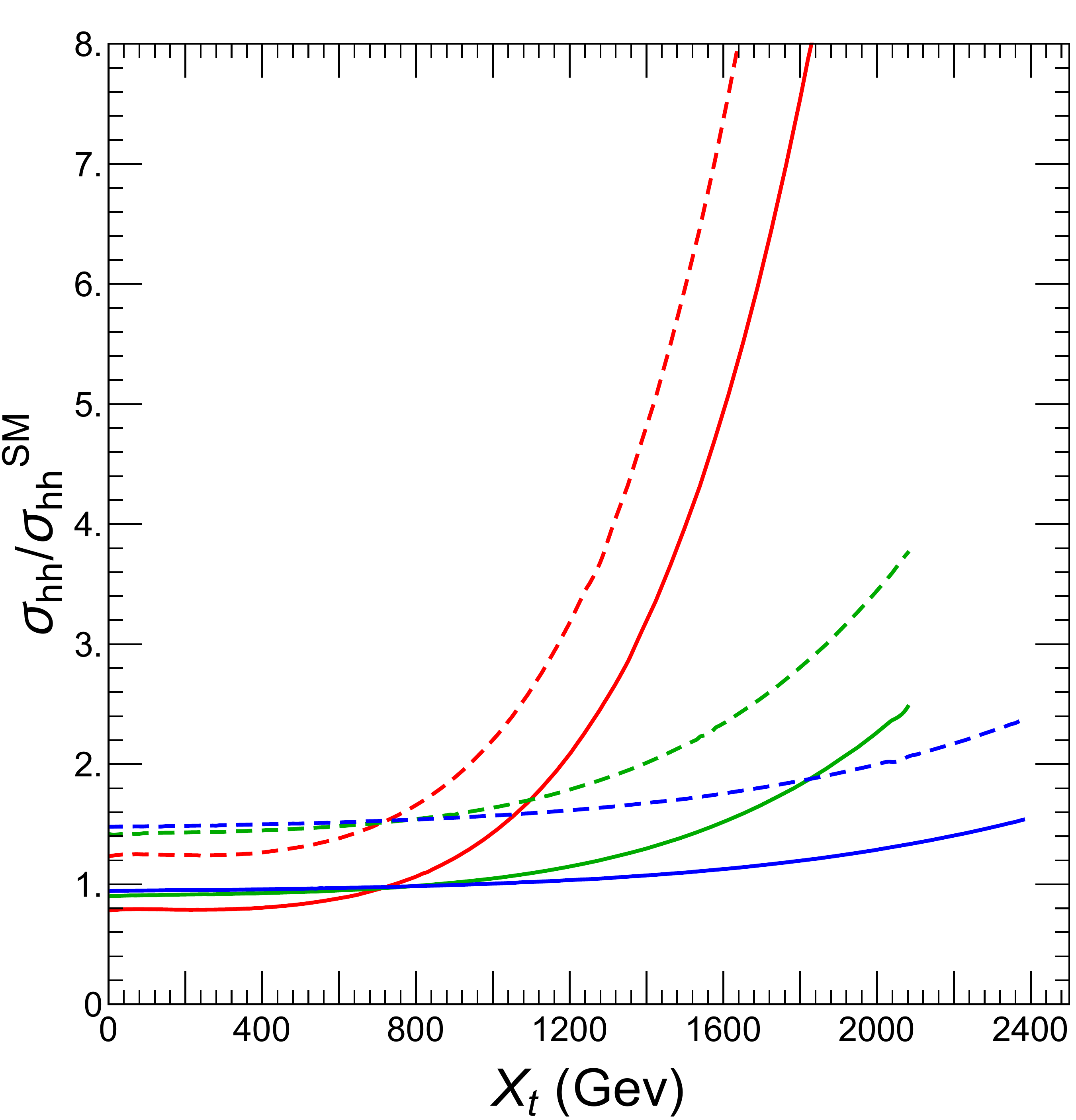}
	\caption{\label{fig:Xtvariance} $X_t$ dependence of the Di-Higgs production cross section for $\delta_3=0$ and $m_Q=m_U$ for lighter stop masses of $300$~GeV (Red), $400$~GeV (Green) and $500$~GeV (Blue) GeV after neglecting the D-terms. Solid lines correspond to $\kappa_t=1$, while dashed lines correspond to $\kappa_t=1.1$. The condition for vacuum stability described in Eq.~(\ref{HgsVcmStb}) is taken to be $X_t<3\,m_Q$, which can always be achieved for a suitably high value of $m_A$ allowed experimentally. For presentational reasons, for the red line is cut at values of $X_t$ smaller than $3\,m_Q$ (see the text).}
\end{figure}

In all cases, when the stops are heavy enough, i.e., above~$\sim$~1 TeV,  the EFT calculation and the full one-loop calculation agree well, as one would expect. Also, with heavy stops, the cross section ratio approaches one for $\kappa_t = 1$ cases, and 1.6 for the $\kappa_t = 1.1$ case (blue line), which is in agreement with the results shown in Fig~\ref{fig:xsecvskt}. The orange line depicts results for $\kappa_g=1$, namely when $X_t^2$ is chosen to be $m_{\tilde{t}_1}^2 + m_{\tilde{t}_2}^2$ (see Eq.~(\ref{kappagmodifiedyt})). In the red and blue lines, instead, $X_t^2$ is chosen to saturate the vacuum stability condition, Eq.~(\ref{HgsVcmStb}), in a conservative way by neglecting the $m_A$ and $m_z$ terms, and the results are in agreement with the ones shown in the  top panels of Figs.~\ref{fig:xsec400},\ref{fig:xsec300} and~\ref{fig:xsec500}, where a similar vacuum stability constraint was considered.  Finally,  the green lines show the dependence of the double Higgs production cross section on the stability bound on $X_t$. In order to show this dependence, we use $m_A$ = 350~GeV, $\mu = 400$~GeV, and $\tan\beta = 1$ to calculate the  saturation of the stability condition, Eq.~(\ref{HgsVcmStb}). Therefore a larger $X_t$ can be allowed, and larger modification to the di-Higgs production cross section can be achieved. For instance, for a lightest stop mass of 500 GeV, the bound on $X_t \simeq 3 \ m_Q$, instead of $X_t \simeq 2.6 \ m_Q$ that is obtained when one neglects the $m_A$ dependence of the  vacuum stability bound, and the modification of the di-Higgs production cross section can be as large as 60~$\%$, compared to about 30~$\%$ when no $m_A$ dependence is considered.

In Fig.~\ref{fig:Xtvariance}, we show the effect of stop mixing parameter  $X_t$ on the di-Higgs production cross section for a fixed value of the mass of the lighter stop in the case of $m_Q=m_U$ and $\delta_3=0$. Red, green and blue represent fixed lighter stop mass of 300, 400 and 500~GeV respectively. Solid lines correspond to $\kappa_t=1$, while dashed lines correspond to $\kappa_t=1.1$. The maximum value of $X_t$ for green and blue lines correspond to the condition $X_t\sim 3m_Q$ as set by taking $m_Q=m_U$ and suitably high experimentally allowed values of $m_A$ in Eq.~(\ref{HgsVcmStb}). The maximum value of $X_t$ used for the red line is less than $3m_Q$ in order to increase the readability of the plot ($X_t < 2.88 m_Q$ and $X_t < 2.78 m_Q$ for the solid and dashed lines, respectively).
	
For lower values of $X_t$ and $\kappa_t = 1$, the di-Higgs production cross section becomes smaller than the SM one and then starts increasing owing to the sign flip for the trilinear coupling of lighter stops  with the Higgs boson, which becomes linearly dependent on $X_t$. 
Therefore, for large values of $X_t$ the contribution to the amplitude that grows quadratically with $X_t$ (last row of Fig.~\ref{fig:MSSM}) becomes dominant and the
cross section grows proportional to $X_t^4$. This behavior is clearly seen in the red line in Fig.~\ref{fig:Xtvariance}, corresponding to a lightest
stop mass of 300~GeV. For the green and blue lines the vacuum stability condition of $X_t<3m_Q$ puts an upper bound on $X_t/m_{\tilde{t}_1}$, and makes
other contribution to the amplitude competitive with those that depend quadratically on $X_t$,  preventing the $X_t^4$ behavior to develop.

As can be seen from Fig.~\ref{fig:Xtvariance}, the increase in $X_t$ leads to significant enhancements in the cross section. For lighter stop masses as low as $300$ GeV, large enhancements by a factor of order ten can be obtained before the vacuum breaking condition $X_t< 3 m_Q$ is met. Even for experimentally more viable  values of the lighter stop mass such as $500$ GeV, considerable enhancements of $60\%$ and $140\%$ are possible for $\kappa_t=1$ and $\kappa_t=1.1$, respectively. 

\subsection{ Di-Higgs Search Channel}

The general strategy in the search for double Higgs is to require one Higgs to decay to a pair of bottoms for enough statistics, as the total rate for double Higgs production is about three orders of magnitude smaller compared to single Higgs production. Then, we can consider the other Higgs decay to a pair of photons, bottoms, $W^{\pm}$'s, or $\tau$'s. In this work, we are going to discuss the modifications to distributions in the presence of light stops, and we will focus on the $bb\gamma\gamma$ channel, as this channel provides the best resolution. 

The cross section for $bb\gamma\gamma$ final state depends not only on the di-Higgs production cross section, but also  on the Higgs decay branching ratios to $bb$ and $\gamma\gamma$. These decay branching ratios depend strongly on the Higgs coupling to $W^\pm$ gauge bosons and bottom quarks, called $\kappa_w$ and $\kappa_b$ respectively. It is then important to see what are the values of $\kappa_w$ and $\kappa_b$ allowed by Higgs data, for the values of $\kappa_t$ and $\kappa_g$ considered in this work. In order to do this, we recall that, the gluon fusion production rate is modified by a factor $\kappa_g^2$, while the vector boson fusion and associated production with vector boson channels are modified by $\kappa_w^2$ and the $t\bar{t}h$ channels are modified by a factor $\kappa_t^2$. Moreover, the modified branching ratios are given by
\begin{align}
BR(h \to X X)&=\frac{ \kappa_X^2 \  BR(h \to XX)^{\rm SM}}{\sum_i \kappa_i^2 \ BR(h \to ii)^{\rm {SM}}}
\end{align}
where $BR(h \to XX)$ is the branching ratio of the Higgs decay into a pair of $X$ particles.

\begin{figure}[tbp]
\centering
\includegraphics[width = .48\textwidth] {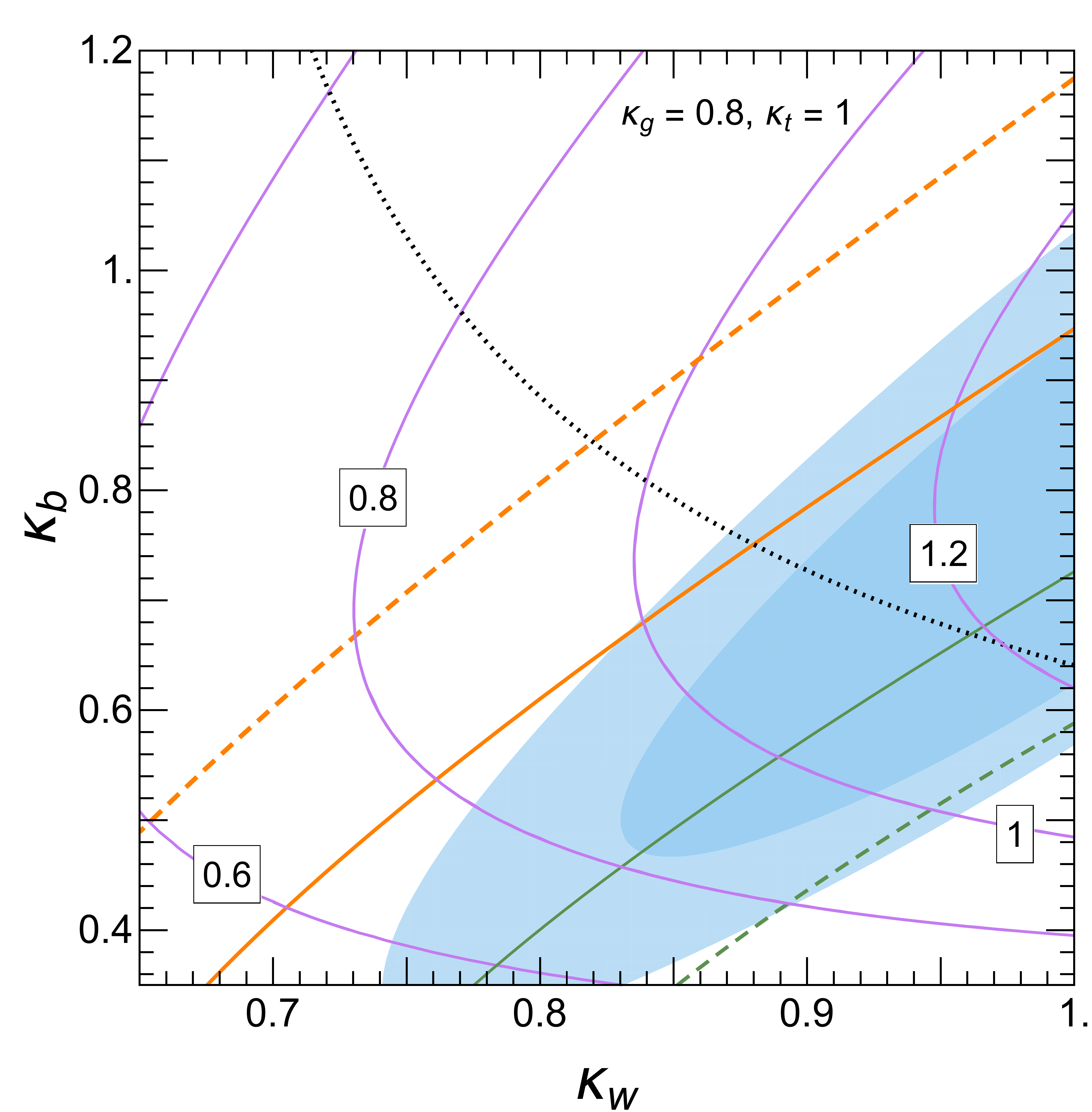}
\hfill
\includegraphics[width = .48\textwidth] {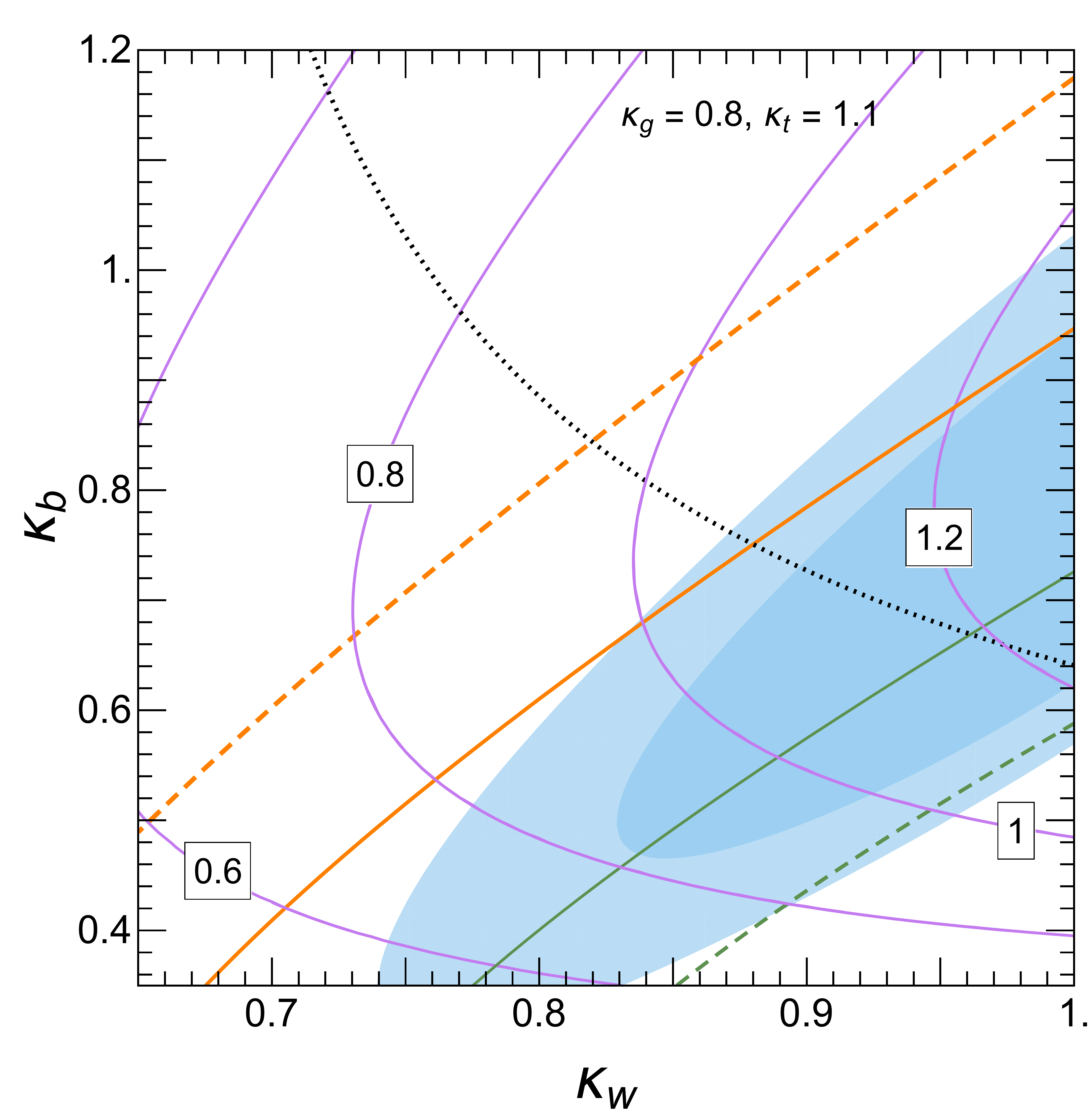}
\hfill
\includegraphics[width = .48\textwidth] {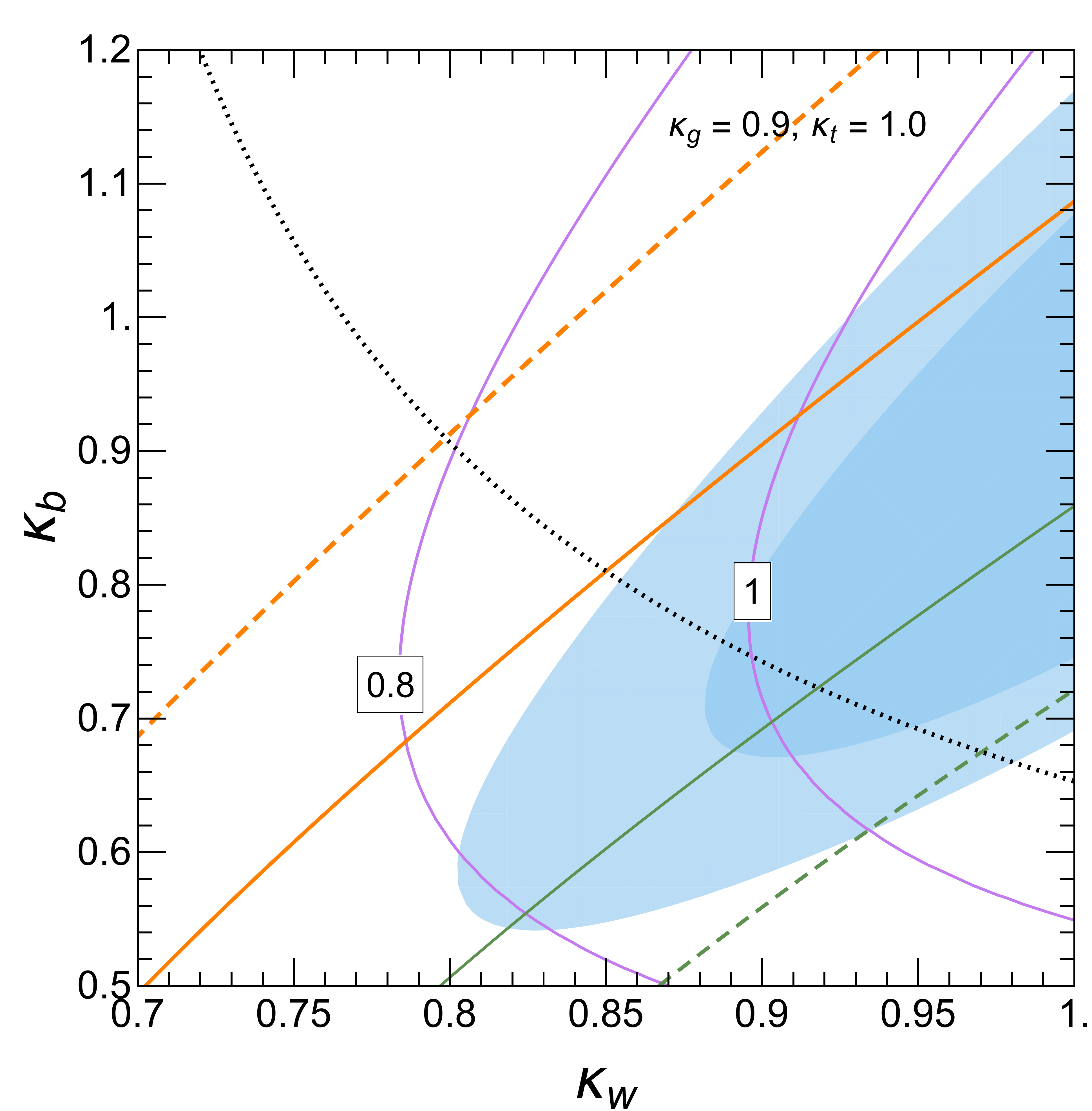}
\hfill
\includegraphics[width = .48\textwidth] {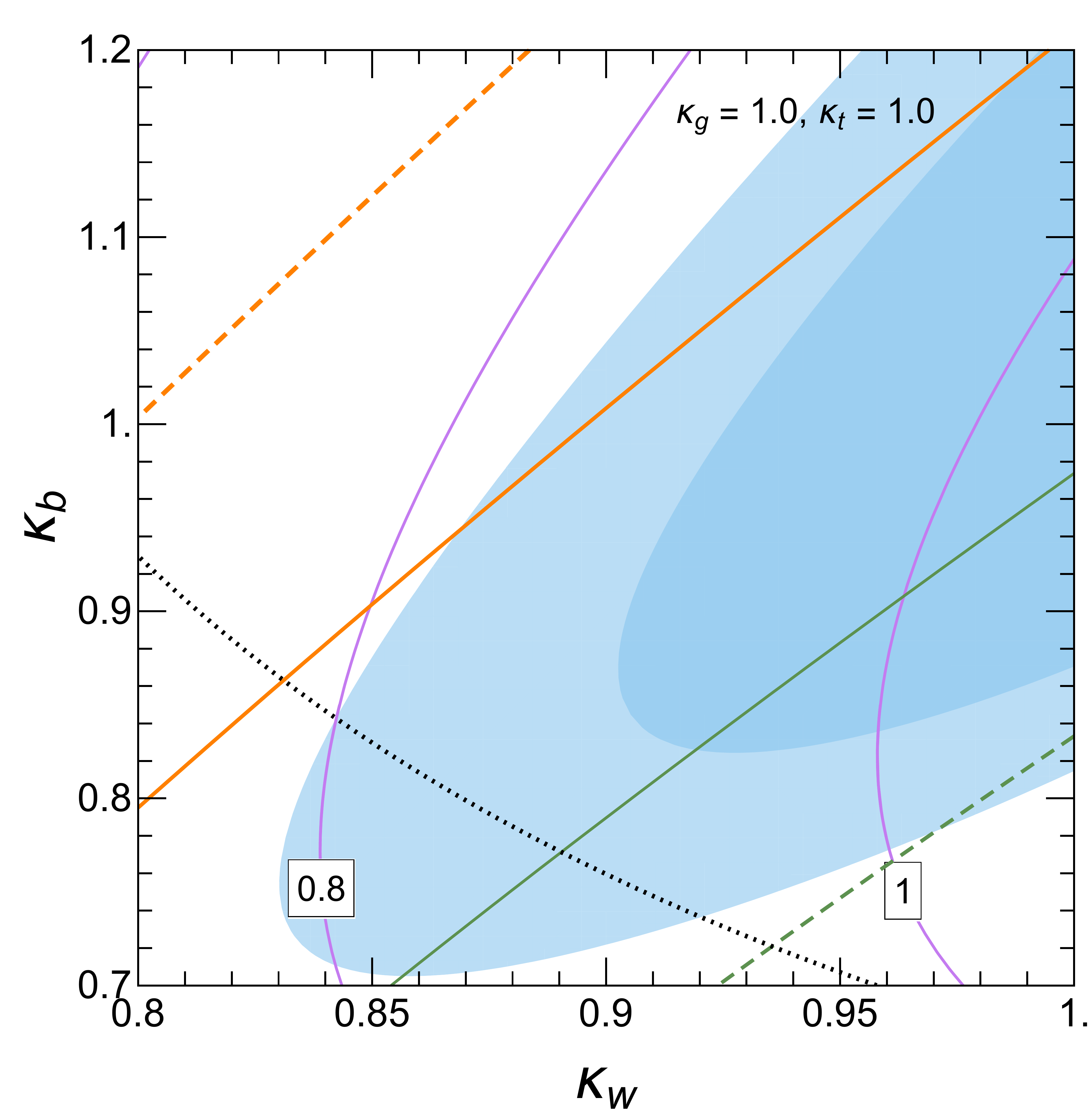}

\caption{\label{fig:BR} The fit for $\kappa_b$ and $\kappa_w$ for the representative values of $\kappa_t$ and $\kappa_g$ we considered earlier in this work. In the top left panel, we present results for $\kappa_g = 0.8$, $\kappa_t = 1$, in the top right panel, for $\kappa_g = 0.8$, $\kappa_t = 1.1$, in the bottom left panel for $\kappa_g = 0.9$, $\kappa_t =1$  and in the bottom right panel for $\kappa_g = 1$, $\kappa_t = 1$. The contours show the values of $BR(h\rightarrow bb) \times BR(h\rightarrow \gamma\gamma)$, normalized to the SM values. Orange and Green contours show the Run-2 gluon fusion $h\rightarrow\gamma\gamma$ results for ATLAS and CMS respectively. The solid lines show the best fit values and the dashed lines show the 1~$\sigma$ lower(ATLAS) and higher(CMS) contours. The region above the black dotted line is consistent with the ATLAS $Vh$, $h\rightarrow b\bar{b}$ measurement within 1~$\sigma$.}
\end{figure}

In Fig.~\ref{fig:BR}, we fix $\kappa_t$ at 1 or 1.1, $\kappa_g$ at  0.80, 0.90 or 1, which are representative values for the gluon Higgs couplings necessary to obtain sizable modifications of the di-Higgs production cross section. Having fixed these values, we fit for the preferred values of  $\kappa_b$ and $\kappa_w$. We include all the Higgs data from Run I~\cite{Khachatryan:2016vau,Aad:2015zhl}, except $h\rightarrow ZZ^*$ and $h\rightarrow \tau\tau$, as they mostly depend on $\kappa_z$ and $\kappa_\tau$, which are beyond the scope of discussion in this study. The production for VBF also depends on $\kappa_z$, which we fix at the run I best fit value $\kappa_z$ = 1. Due to the small value of the $BR(h \to ZZ)$, fixing $\kappa_z = \kappa_w$ makes no difference in our results. The value of $\kappa_\gamma$ is considered to be consistent with the values induced by the presence of light stops and modifications of $\kappa_t$ and $\kappa_w$. Using 
effective field theory to evaluate the top and stop contributions, one obtains, approximately
\begin{equation}
\kappa_{\gamma} = 1.28 \kappa_w - 0.28 \kappa_g,
\end{equation}
where we used Eq.~(\ref{kappagmodifiedyt}) and the fact that the relation between the top and stop contributions to $\kappa_g$ and $\kappa_\gamma$ 
are the same.

The region within 1~$\sigma$ of the best fit value for $\kappa_b$ and $\kappa_w$ is shown in blue, and the region within 2~$\sigma$ of the best fit value is shown in light blue. Then, for given values of $\kappa_w$ and $\kappa_b$, we calculate the Higgs decay branching ratios to $bb$ and $\gamma\gamma$, and we show the contours of $BR(h\rightarrow bb) \times BR(h\rightarrow \gamma\gamma)$, normalized to the SM value.  We also show the Run 2 results for gluon fusion, $h\rightarrow \gamma \gamma$ in orange(ATLAS)~\cite{ATLAS:2017myr}, and green(CMS)~\cite{CMS:2017rli}. The solid lines are the central values, and the dashed line show the 1~$\sigma$ range. The region above the dotted line is consistent with the Run 2 measurement of associated production of Higgs with vector bosons, $Vh$, with $h\rightarrow b\bar{b}$ within 1~$\sigma$~\cite{Aaboud:2017xsd}. It can be seen from the top two panels that $\kappa_t$ does not change the fit, as the tth channel has large uncertainties. $\kappa_t$ does not change the branching ratios either, because by allowing new particles in the loop, or considering $\kappa_g$ as an independent parameter, $\kappa_t$ does not change the Higgs decay. Then for $\kappa_g = 0.9$ and $\kappa_g = 1$ we only consider $\kappa_t = 1$.

Our results are roughly consistent with the ones obtained by the combined ATLAS and CMS Higgs data~\cite{Khachatryan:2016vau,Aad:2015zhl}. 
As can be seen from  these contours, some small modifications to $BR(h\rightarrow bb) \times BR(h\rightarrow \gamma\gamma)$ are expected, which would modify the $hh\rightarrow bb\gamma\gamma$ rate. However, the largest modification is about $\pm20\%$. Let us stress that the inclusion of run II data is likely to move $\kappa_b$ towards larger values.  However, as is apparent from Fig.~\ref{fig:BR}, this modification is unlikely to modify the
above conclusion. Therefore, only mild variations are expected in the product of the $bb$ and $\gamma \gamma$ decay branching ratios and
the $hh\rightarrow bb \gamma\gamma$ rate is mainly controlled by the modifications of the di-Higgs production rate with respect to the SM value. 

\subsection{Modifications of the di-Higgs invariant mass distribution}

As pointed out in ~\cite{Barger:2013jfa,Huang:2015tdv}, a modification in $\lambda_3$ can lead to a drastic change in the kinematic distributions for double Higgs production. The $m_{hh}$ distribution shifts significantly to  lower values for $\delta_3 \gtrsim 2$. In this section, we study the possible modifications to the $m_{hh}$ distribution with modified $\kappa_t$ and the presence of a light stop. 

\begin{figure}[tbp]
\centering
\includegraphics[width = .75\textwidth]{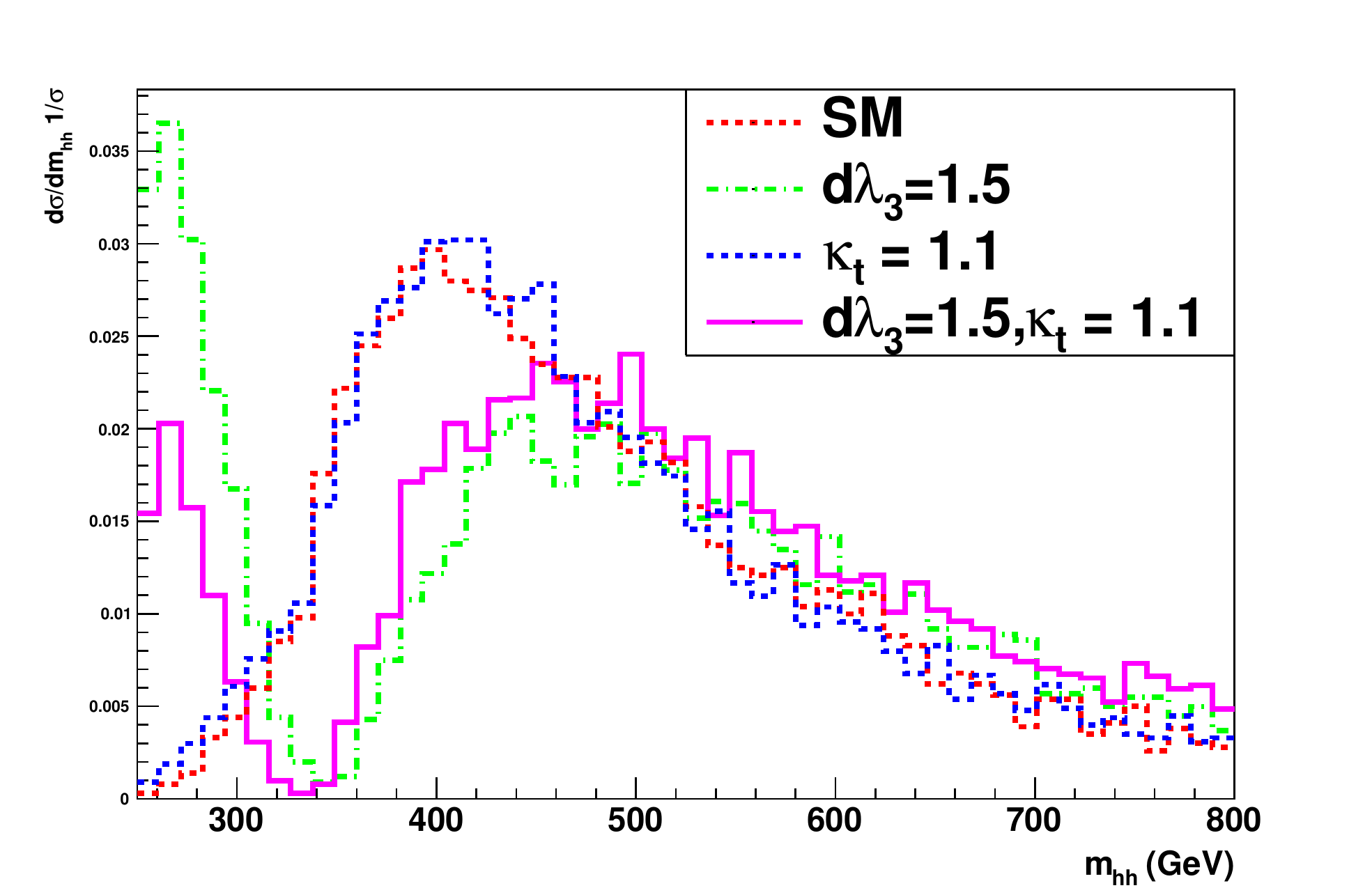}
\caption{\label{fig:nostop} Normalized $m_{hh}$ distribution with modified $\lambda_3$, and $y_t$.}
\end{figure}

As we emphasized before,
in the SM, there are two diagrams contributing to the double Higgs production, the box diagram and the triangle diagram. The two diagrams interfere with each other destructively. The cross section for the triangle diagram scales with $\kappa_t$ as $\kappa_t^2$, and the cross section of the box diagram scales with $\kappa_t$ as $\kappa_t^4$. Therefore, a modification in the top Yukawa can change not only the di-Higgs production cross section but also the $m_{hh}$ distribution. However, without modifications in $\lambda_3$, the box diagram dominates over the triangle diagram, and as only a few tens of percent deviation is allowed in the top Yukawa, we do not expect that this change can modify the $m_{hh}$ distributions in any relevant way, as we have checked in our numerical simulations and can be seen from the blue dashed line in Fig~\ref{fig:nostop}. 

The modification of the invariant mass distribution could become relevant when the cancellation between the two diagrams become strong. This
occurs for values of  $\lambda_3 \sim 2.5$, for which the amplitudes associated with the two diagrams become comparable in size.  In this case, 
a cancellation of the production rate appears at some value of $m_{hh}$. 
 Then modifications of $\kappa_t$ would change  the relative weight of the triangle and box diagrams, inducing a more relevant change in the invariant mass distribution, $m_{hh}$.
This can be seen from the green dot dashed line and the magenta solid line in Fig.~\ref{fig:nostop}, where $\lambda_3$ is 2.5$ \lambda_3^{SM}$ in both lines. When $\kappa_t$ is 1 (green dot dashed line), the cancellation is at $m_{hh}$ about $2 m_{t}$. As $\kappa_t$ increases to 1.1 (magenta solid line), the box diagram increases more than the triangle diagram, and the exact cancellation occurs for smaller values of  $m_{hh}$, at about 330~GeV.  

\begin{figure}[tbp]
 \centering
\includegraphics[width = .75\textwidth]{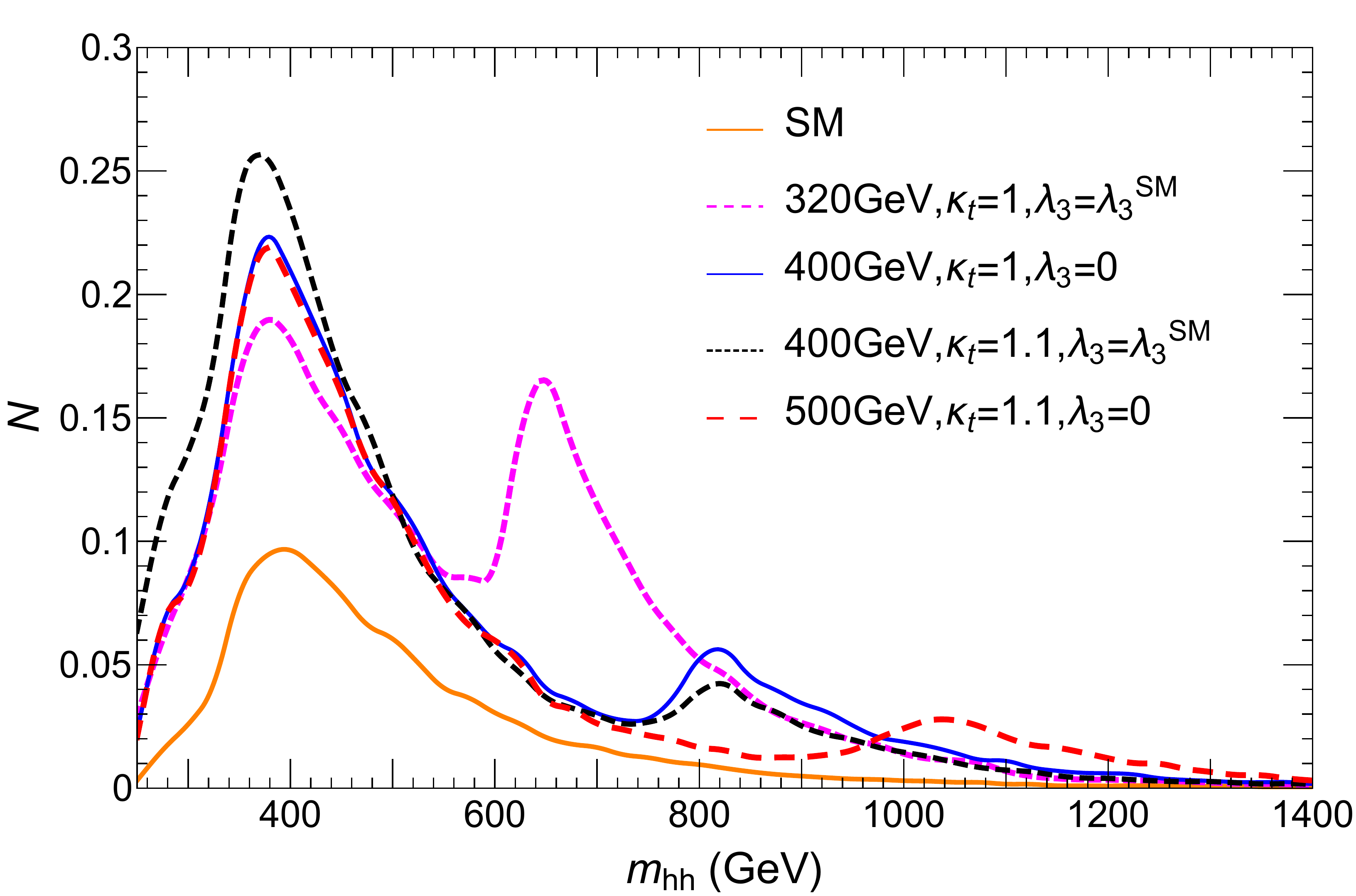}
\caption{\label{fig:mhhstop} $m_{hh}$ distribution in presence of light stops, possible modifications in $\lambda_3$ and $y_t$. BMs are described in Table~\ref{tab:bm}. The distributions are normalized to the number of events for 3 $ab^{-1}$ integrated luminosity at HL-LHC.}
\end{figure}

Furthermore, in the presence of a light stop, the amplitudes for diagram (3) - (8) in Fig.~\ref{fig:MSSM}
develop imaginary parts when the invariant mass $m_{hh}$ crosses the 2~$m_{\tilde{t}}$ threshold,  inducing a second peak in the $m_{hh}$ distribution a little above 2~$m_{\tilde{t}}$. We selected  benchmarks to study the distributions of $m_{hh}$ with a light stop, and possible modifications in $\lambda_3$ and $\kappa_t$. The benchmarks are listed in Table~\ref{tab:bm}, in which we are neglecting D-terms when calculating the stop mass. This effect can be seen in Fig~\ref{fig:mhhstop}. The benchmark points B and C have a light stop around 400 GeV, so there is a second peak around 800~GeV, while 
benchmark points A and D corresponds to stop masses of about 320~GeV and 500~GeV, respectively, and therefore the second peak is around 640~GeV and 1~TeV, respectively.  
\begin{table}
\centering
\begin{tabular}{|c|c|c|c|c|c|c|c|}
\hline
 & $m_{U,Q}$  (GeV) &  $X_t $ (GeV) & $\frac{\lambda_3} {\lambda_3^{SM}}$ &$\kappa_t$ & $\kappa_g$& $m_{\tilde{t}_1}$ (GeV) & $\frac{\sigma_{hh}}{\sigma_{hh}^{SM}}$\\
 \hline
 BMA & 575 & 1495 & 1 & 1 & 0.82 & 320 &2.87 \\
 \hline
 BMB & 650 & 1690 & 0 & 1 & 0.88 & 400 & 2.73 \\
 \hline
 BMC & 650 & 1690 & 1 & 1.1 & 0.96 & 400 & 2.53\\
 \hline  
 BMD & 795 & 2385 & 0 & 1.1 & 0.97 & 500 &  2.4\\
 \hline
 \end{tabular}
 \caption{ \label{tab:bm} Benchmarks points for light stops giving a sizable correction to the di-Higgs production cross section at hadron colliders}
 \end{table}

As can be seen from Fig.~\ref{fig:mhhstop}, the kinematic distributions are similar to the ones in the SM and hence the $m_{hh}$ cut efficiency in this benchmark points is similar to the SM case. Other kinematic variables that have been used at the LHC, including the invariant mass distributions of the bottom quarks, $m_{bb}$,  and of the diphotons, $m_{\gamma\gamma}$, as well as cuts on the $p_T$ of the b-jets, and the photons are expected to have a similar behavior as in the SM. Therefore, the projected sensitivity scales approximately with the signal rate and therefore we use the ATLAS  SM results to estimate the projected sensitivity for our benchmarks at the  High Luminosity run of the LHC (HL-LHC)~\cite{ATL-PHYS-PUB-2017-001}, with a projected luminosity of 3~$ab^{-1}$, in Table~\ref{tab:LHC}. CMS shows a similar sensitivity in this channel~\cite{CMS:2017ihs}. 

A recent work proposes to use the log-likelihood ratio to identify kinematic regions and shows an improved sensitivity~\cite{Kling:2016lay}. As one can see from Table~\ref{tab:LHC}, only using the $bb\gamma\gamma$ channel, the HL-LHC will be sensitive to light stops with a large mixing, which can be a indirect probe for light stops regardless how the stops may decay. For stops as heavy as 500 GeV, the LHC sensitivity is limited to the cases of a large mixing,  a negative correction to the Higgs trilinear coupling, which is well motivated by a strong first order phase transition, and/or a small positive correction to the top-quark Higgs coupling, such as it appears in benchmark point D.

\begin{table}
\centering
\begin{tabular}{|c|c|c|c|c|c|}
\hline 
$S$/$\sqrt{B}$ & SM & BMA & BMB & BMC & BMD \\
\hline 
 & 1.1 & 3.0 & 2.9 & 2.7 & 2.5 \\
 \hline
\end{tabular}

\caption{\label{tab:LHC} Projected sensitivities for the benchmarks points at the HL-LHC, using only the $bb\gamma\gamma$ channel.}
\end{table}

In summary, the presence of a light stop, modifications of the top Yukawa and trilinear Higgs couplings can lead to sizable contribution to double Higgs production. The stop contributions are summarized in Fig.~\ref{fig:EFTcomparison}, the contributions from a modified top Yukawa coupling is summarized in Figs.~\ref{fig:xsecvskt}. We present some benchmarks and their projected sensitivities in Table~\ref{tab:bm} and Table~\ref{tab:LHC}.  
\section{Conclusions}
The search for  di-Higgs production is one of the main goals at hadron colliders. This is due to the sensitivity of this channel to new physics and its dependence on the Higgs potential parameters. The sensitivity of the LHC experiments to this channel is limited by the small rate and large backgrounds in the main final state channels. It is therefore very important to study under which conditions the di-Higgs production rate may be enhanced, allowing for its study at a high luminosity LHC. Barring the possibility of resonance di-Higgs production via the presence of heavy scalars decaying into pairs of SM-like Higgs bosons, it is known that this can be done in the presence of negative corrections to the trilinear Higgs coupling and/or positive corrections to the top quark coupling to the Higgs. In this work we emphasized the strong dependence of the di-Higgs production cross section to small, positive corrections of the top-quark coupling to the Higgs, which are still allowed by the current LHC Higgs data.

Furthermore, we studied the additional effects of light stops on the di-Higgs production cross section. We computed the one-loop corrections associated with light stops, finding agreement with previous expressions in the literature. We then incorporated these corrections into a modified version of the program MCFM-8.0, including the possibility of light stops together with possible modifications of the top-quark and trilinear Higgs couplings.  We found out that large corrections to the di-Higgs production rate are possible in the case of relatively light stops, with a large stop mixing parameter.  The effect of light stops may become even stronger under  modifications of the top-quark or trilinear Higgs couplings. In general, we found that the modifications of the di-Higgs production rate are strongly correlated with similar modifications of the gluon fusion Higgs production rate,  and can significantly enhance the LHC sensitivity to this production channel. Moreover, we  also found that the precise constraints on the trilinear Higgs-stop coupling coming from the requirement of vacuum stability, have a major impact on the size of the possible stop corrections for lightest stop masses above 500~GeV, that is the current stop bound on the stop mass in standard decay channels.

\acknowledgments
Work at University of Chicago is supported in part by U.S. Department of Energy grant number DE-FG02-13ER41958. Work at ANL is supported in part by the U.S. Department of Energy under Contract No. DE-AC02-06CH11357. Work at Texas A$\&$M is supported in part by U.S. Department of Energy grant number DE-SC0010813. Work at University of Nebraska-Lincoln is supported by the Department of Physics and Astronomy. The work of C.W. was partially performed at the Aspen Center for Physics, which is supported by National Science Foundation grant PHY-1607611. P.H. thanks the Mitchell Institute for Fundamental Physics and Astronomy for support. Part of the work by A.J. was performed as a visitor at Amherst Center for Fundamental Interactions (ACFI) at University of Massachusetts Amherst, which was supported under U.S. Department of Energy Contract No. DE-SC0011095. A.J. thanks Michael Ramsey-Musolf for the hospitality and support at ACFI. We would like to thank John Campbell for help in handling the MCFM program and Brian Batell for useful discussions and comments. 

\newpage
\appendix
\allowdisplaybreaks
\section*{Appendix}

	\section{Form Factors  }
	\label{appendix1}
	In this section, we present scalar form factors we use.  The cross section of $gg\to hh$ is determined by two Lorentz and gauge invariant structure functions given in in~\cite{'tHooft:1978xw,Asakawa:2010xj,Kribs:2012kz,Glover:1987nx,Belyaev:1999mx,BarrientosBendezu:2001di}. The differential cross section reads,
	
	\begin{eqnarray}\label{SM_di-higgs}
	\frac{\ud \sigma_{gg \to hh}}{\ud \hat{t}} = \frac{|\mathcal{M}|^2}{16\pi\hat{s}^2}
	\end{eqnarray}
	Here the matrix-element $\M$ are separate by four parts according to the helicity of the incoming glouns,  where $+$ and $-$ denotes the right- and left helicity gluons.  Thus~\cite{Bartl:1997yd}:
	
	\begin{eqnarray}
	\Lvert\M\Lvert^2=\frac{\alpha_S^2}{2^{10}}\Big[\Lvert \sum_{i,j=1}^{2}M_{++}^{(n)}(\tilde{t}_i,\tilde{t}_j)\Lvert^2+\Lvert \sum_{i,j=1}^{2}M_{+-}^{(n)}(\tilde{t}_i,\tilde{t}_j)\Lvert^2\Big]
	\label{eq:amplitude1}
	\end{eqnarray}
     The amplitude can also be written in terms of form factors of triangle $F_{\triangle}$, and box diagrams, $ F_{\Box},\; G_{\Box}$, with trilinear coupling $C_{\triangle}=\frac{3m_h^2}{\hat{s}-m_h^2}$ and quartic coupling $C_{\Box}$ normalized to unity \cite{Plehn:1996wb},
     
     \begin{eqnarray}\label{form factor}
     |\mathcal{M}|^2=\frac{\alpha_s^2G_F^2\hat{s}^2}{2^{11}\pi^2}\big\{|F_{\triangle}C_{\triangle}+F_{\Box}C_{\Box}|^2+|G_{\Box}C_{\Box}|^2\big\},
  \label{eq:amplitude2}
     \end{eqnarray}
    where the two terms inside the bracket in Eq.~(\ref{eq:amplitude2}) are in one to one correspondence with the ones given in  Eq.~(\ref{eq:amplitude1}) and $G_F=\frac{1}{\sqrt{2}v^2}$ is the Fermi constant.  We will discuss the value of the above form factors below.
	
	The above amplitudes depend on  the couplings $g_{hab}$ of the mass eigenstates of quarks and squarks to the Higgs field.  The stop mass eigenstate couplings may be obtained from the corresponding coupling of the weak eigenstates, namely
	\begin{eqnarray}
	\mathbf{G}_{h\tilde{t}\tilde{t}} =\mathbf{R}\cdot
	\left( \begin{array}{cc}
	g_{h\tilde{t}_L^{*}\tilde{t}_L} & g_{h\tilde{t}^*_L\tilde{t}_R} \\
	g_{h\tilde{t}^*_L\tilde{t}_R} & g_{h\tilde{t}_R^{*}\tilde{t}_R} \\
	\end{array} \right)\cdot\mathbf{R}^{\mathbf{T}},
	\end{eqnarray}
where $\mathbf{R}$ is the rotational matrix which rotates the left- and right- handed squark fields to the mass eigenstates.~\cite{BarrientosBendezu:1999gp}.  The mixing angle $\theta$ of the $2\times 2$ rotation matrix :
 \begin{eqnarray}
 	\sin2\theta_{\tilde{t}}=\frac{\sqrt{2}y_t vX_t}{m_{\tilde{t}_2}^2-m_{\tilde{t}_1}^2}\qquad
 	\cos2\theta_{\tilde{t}}=\frac{m_Q^2-m_U^2}{m_{\tilde{t}_2}^2-m_{\tilde{t}_1}^2}\qquad
 \end{eqnarray}	

Considering only small deviations of the Higgs couplings to fermions, close to the decoupling limit $\alpha \simeq \beta-\pi/2$ and neglecting the small contribution of the $D$ terms, the linear and bilinear couplings of the stop mass eigenstate to the Higgs are given by
	\begin{eqnarray}\label{StpCpling}
	g_{htt}&=&-\kappa_t\frac{m_t}{v}\\
	g_{h\tilde{t}_{1,2}\tilde{t}_{1,2}}&=&-2\kappa_t\frac{m_t^2}{v}(1\mp\frac{X_t}{2m_t}
	\sin{2\theta_{\tilde{t}}})\\
	g_{h\tilde{t}_1\tilde{t}_2}&=&\kappa_t\frac{m_t}{v}X_t
	\cos 2\theta_{\tilde{t}}\\
	g_{hh\tilde{t}_i\tilde{t}_i}&=&-2\kappa_t^2(\frac{m_t}{v})^2
	\end{eqnarray}
	The form factors are associated with the  diagrams given in Fig.~\ref{fig:MSSM}, where the Higgs triple coupling constant $\lambda_3=-(1+\delta_3)\frac{3m_h^2}{v}$
	\begin{align}
	M_{++}^{(1)}&=-\frac{m_tg_{htt}\lambda_3}{\pi[\hat{s}-m^2_h]}\Big[2+(4m_t^2-\hat{s})C_{ttt}^{00}(\hat{s})\Big]\\
	M_{+-}^{(1)}&=0\\
	M_{++}^{(2)}&=\frac{g_{htt}g_{htt}}{2\pi\hat{s}}\Big\{-4\hat{s}-8m_t^2C_{ttt}^{00}(\hat{s})\hat{s}-2\Big(4m_t^2-m_h^2\Big)\Big[2TC_{ttt}^{h0}(\hat{t})+2UC_{ttt}^{h0}(\hat{u})\nonumber\\
	&-(m_h^4-\hat{t}\hat{u})D_{tttt}^{h0h0}(\hat{t},\hat{u})\Big]-2m_t^2(8m_t^2-2m_h^2-\hat{s})\hat{s}\nonumber\\
	&\times\Big[D_{tttt}^{h0h0}(\hat{t},\hat{u})+D_{tttt}^{hh00}(\hat{s},\hat{t})+D_{tttt}^{hh00}(\hat{s},\hat{u})\Big]\Big\}\\
	M_{+-}^{(2)}&=-\frac{g_{htt}g_{htt}}{2\pi[m_h^4-\hat{t}\hat{u}]}\Big\{(8m_t^2-\hat{t}-\hat{u})(2m_h^4-\hat{t}^2-\hat{u}^2)C_{ttt}^{hh}(\hat{s})\nonumber\\
	&+(m_h^4-8m_t^2\hat{t}+\hat{t}^2)\Big[2TC_{ttt}^{h0}(\hat{t})-\hat{s}C_{ttt}^{00}(\hat{s})+\hat{s}\hat{t}D_{tttt}^{hh00}(\hat{s},\hat{t})\Big]\nonumber\\
	&+(m_h^4-8m_t^2\hat{u}+\hat{u}^2)\Big[2UC_{ttt}^{h0}(\hat{u})-\hat{s}C_{ttt}^{00}(\hat{s})+\hat{s}\hat{u}D_{tttt}^{hh00}(\hat{s},\hat{u})\Big]\nonumber\\
	&+2m_t^2(m_h^4-\hat{t}\hat{u})(8m_t^2-\hat{t}\hat{u})\Big[D_{tttt}^{h0h0}(\hat{t},\hat{u})+D_{tttt}^{hh00}(\hat{s},\hat{t})+D_{tttt}^{hh00}(\hat{s},\hat{u})\Big]\Big\}\\	\label{stpfmfctr}
	M_{++}^{(3+4)}&=\frac{g_{h\tilde{t}_i\tilde{t}_i}\lambda_3}{2\pi(\hat{s}-m_h^2)}\bigg(1+2m_{\tilde{i}_i}C_{iii}^{00}(\hat{s})\bigg)\\
	M_{+-}^{(3+4)}&=0\\
	M_{++}^{(5+6)}&=-\frac{g_{\tilde{t}_i\tilde{t}_ihh}}{2\pi} \bigg(1+2m_{\tilde{t}_i}^2C_{iii}^{00}(\hat{s})\bigg)\\\label{M++(56)}
	M_{+-}^{(5+6)}&=0\\\label{M+-(7)}
	M_{++}^{(7)}&=-\frac{1}{\pi} g_{h\tilde{t}_i\tilde{t}_j}g_{h\tilde{t}_i\tilde{t}_j}C_{iji}^{hh}(\hat{s})\\\label{M++(8)}
	M_{+-}^{(7)}&=0\\
	M_{++}^{(8)}&=\frac{g_{h\tilde{t}_i\tilde{t}j}g_{h\tilde{t}_i\tilde{t}j}}{2\pi\hat{s}}\nonumber\\
	&\times\bigg\{T\big(C_{ijj}^{h0}(\hat{t})+C_{jii}^{h0}(\hat{t})\big)+U\big(C_{ijj}^{h0}(\hat{u})+C_{jii}^{h0}(\hat{u})\big)+2\hat{s}C_{iji}^{hh}(\hat{s}) 
	\nonumber\\
	&+\big[(m_{\tilde{t}_i}-m_{\tilde{t}_j})\hat{s}-(m_h^4-\hat{t}\hat{u})\big]D_{ijji}^{h0h0}(\hat{t},\hat{u}) \nonumber\\&+2\hat{s}m_{\tilde{t}_j}^2\bigg[D_{ijji}^{h0h0}(\hat{t},\hat{u})+D_{jijj}^{hh00}(\hat{s},\hat{t})+D_{jijj}^{hh00}(\hat{s},\hat{u})\bigg]\bigg\}\\
	M_{+-}^{(8)}&=\frac{g_{h\tilde{t}_i\tilde{t}_j}g_{h\tilde{t}_i\tilde{t}_j}}{2\pi(m_h^4-\hat{t}\hat{u})}\bigg\{\hat{s}\big(2(m_{\tilde{t}_i}^2-m_{\tilde{t}_j}^2)+\hat{t}+\hat{u}\big)C_{iii}^{00}(\hat{s})-2\hat{t}TC_{jii}^{h0}(\hat{t})-2\hat{u}UC_{jii}^{h0}(\hat{u})\nonumber\\
	&-T(m_{\tilde{t}_i}^2-m_{\tilde{t}_j}^2)\big[C_{ijj}^{h0}(\hat{t})+C_{jii}^{h0}(\hat{t})\big]-U(m_{\tilde{t}_i}^2-m_{\tilde{t}_j}^2)\big[C_{ijj}^{h0}(\hat{u})+C_{jii}^{h0}(\hat{u})\big]
	\nonumber\\
	&+\Big(2m_h^4-\hat{t}^2-\hat{u}^2\Big)C_{iji}^{hh}(\hat{s})+\Big[-\hat{s}(m_{\tilde{t}_i}^2-m_{\tilde{t}_j}^2)^2+(m_{\tilde{t}_i}^2+m_{\tilde{t}_j}^2)(m_h^4-\hat{t}\hat{u})\Big]\nonumber\\
	&\quad\times\Big[D_{jijj}^{h0h0}(
	\hat{t},\hat{u})+D_{ijii}^{hh00}(\hat{s},\hat{t})+D_{ijii}^{hh00}(\hat{s},\hat{u})\Big]\nonumber\\
	&+\Big[-\hat{s}\hat{t}^2-(m_{\tilde{t}_i}^2-m_{\tilde{t}_j}^2)\Big(2\hat{t}\hat{s}-(m_{h}^4-\hat{t}\hat{u})\Big)\Big]D_{ijii}^{hh00}(\hat{s},\hat{t})\nonumber\\
	&+\Big[-\hat{s}\hat{u}^2-(m_{\tilde{t}_i}^2-m_{\tilde{t}_j}^2)(2\hat{u}\hat{s}-(m_h^4-\hat{t}\hat{u}))\Big]D_{ijii}^{hh00}(\hat{s},\hat{u})\bigg\}\label{eqn:A19}
	\end{align}
	
	Here $p_1, \;p_2$ are the momentum of incoming glouns, $k_1,\;k_2$ outgoing Higgs, with $p_1^2=p_2^2=0$, $k_1^2=k_2^2=m_h^2$.  $U=(m_h^2-\hat{u}),\,T=(m^2_h-\hat{t})$ and $S=(m_h^2-\hat{s})$. $C_{ijk}^{ab}(\hat{\alpha})$ and $D_{ijkl}^{abcd}(\hat{\alpha},\hat{\beta})$ are defined in terms of the Passarino-Veltman functions C and D, which are given in~\cite{Passarino:1978jh}
	\begin{eqnarray}
	C_{ijk}^{ab}(\hat{\alpha})&=&C(m_a^2,m_b^2,\hat{\alpha},m_i^2,m_j^2,m_k^2)\\
	D_{ijkl}^{abcd}(\hat{\alpha},\hat{\beta})&=&D(m_{a}^2,m_{b}^2,m_{c}^2,m_d^2,\hat{\alpha},\hat{\beta},m_i^2,m_j^2,m_k^2,m_l^2),
	\end{eqnarray}
	and $\hat{s}$, $\hat{u}$ and $\hat{t}$ are the standard partonic Mandelstam variables for this process, $m_{i}= m_{\tilde{t}_i}$. 
\section{Effective Field Theory Analysis}
	\label{EFTanalysis}
\begin{figure}[tbp]
		\centering
		\includegraphics[width = 0.9\textwidth]{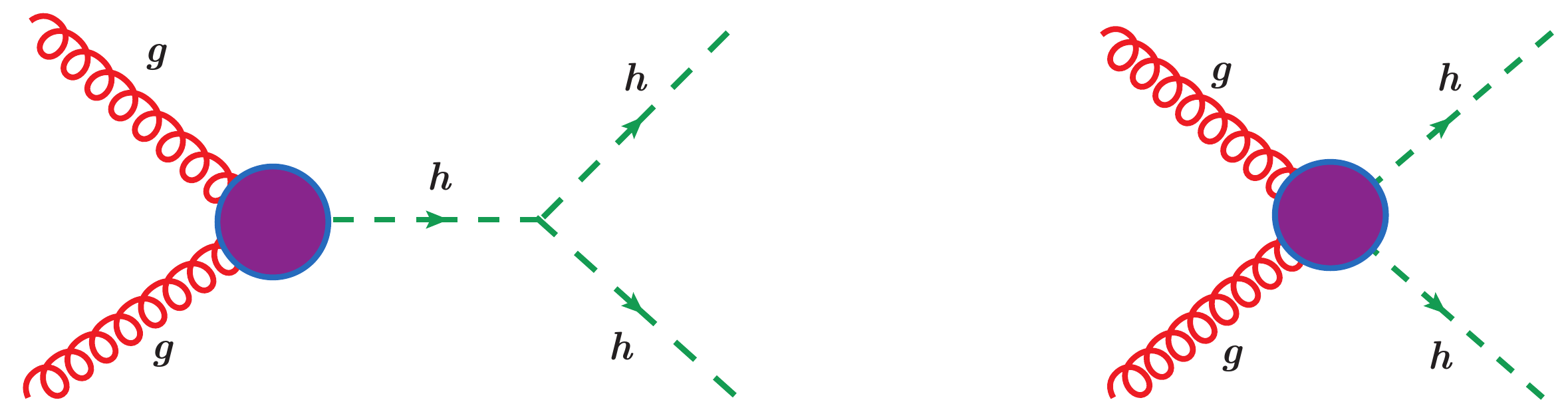}
		\caption{\label{fig:gg-hh_EFT} $gg\rightarrow hh$ with EFT vertices.}
	\end{figure}

 	In the limit that the colored particles in QCD loop is much heavier than the relevant energy scale in the theory, $m_Q\gg m_{hh}$, the form factors in (\ref{form factor}) may be computed using effective field theory techniques for the effective vertices shown in Fig.~\ref{fig:gg-hh_EFT}. In Sec.~\ref{sec:xsec}, we discussed the leading contribution to the single Higgs production. For the di-Higgs production, 
	according to Eq.(\ref{EFT_Taylor_expansion}), the second order coupling of the Higgs, necessary for the computation of the  $gg\rightarrow hh$
	amplitude, can be written as:
	\begin{eqnarray}\label{4}
	\mathcal{L}_{gg\rightarrow hh}= & - & \frac{\alpha_{s}}{16\pi}G^{a}_{\mu\nu}G^{a\mu\nu}\left(\frac{h}{\upsilon}\right)^{2}\sum_{i}\beta_{i}\frac{\partial\log [\det{M^{\dagger}(\upsilon)M(\upsilon)}]}{\partial(\log\upsilon)} \nonumber\\
	& + &\frac{\alpha_{s}}{16\pi}G^{a}_{\mu\nu}G^{a\mu\nu}\left(\frac{h}{\upsilon}\right)^{2}\sum_{i}\beta_{i}\frac{\partial^{2}\log [\det{M^{\dagger}(\upsilon)M(\upsilon)}]}{\partial(\log\upsilon)^{2}}
	\end{eqnarray}
We can now  define dimensionless couplings $g_{h}^{(i)}=\frac{\partial\log [\det{M^{\dagger}M}]}{\partial\log\upsilon}$ and $g_{hh}^{(i)}=\frac{\partial^{2}\log [\det{M^{\dagger}M}]}{\partial(\log\upsilon)^{2}}$.  
This may be identified with the form factors defined in Eq.~(\ref{form factor})
\begin{eqnarray}
	F_{\triangle} & = & \frac{1}{8}\beta_i g_h^i\\
	F_{\Box} &= & \frac{1}{8}\beta_i(-g_h^i+g_{hh}^i)\\
	G_{\Box}& =& 0, 
	\end{eqnarray}
In the limit of vanishing soft supersymmetry breaking terms, these form factors become constants, proportional to the particle contributions
to the QCD $\beta$ function \cite{Dawson:2015oha},
	\begin{eqnarray}\label{lrgmslmt}
	F_{\triangle}^f &\sim& \frac{2}{3}=\frac{1}{2}\beta_{f}^{\rm Dirac} \qquad
	F_{\Box}^f \sim -\frac{2}{3}=-\frac{1}{2}\beta_{f}^{\rm Dirac}\qquad\qquad\nonumber\\
	F_{\triangle}^s &\sim& \frac{1}{6}=\frac{1}{2}\beta_{0}^{\rm scalar}\qquad
	F_{\Box}^s \sim -\frac{1}{6}=-\frac{1}{2}\beta_{0}^{\rm scalar}\qquad\nonumber\\
	G_{\Box}&\sim& \mathcal{O}(\frac{P_{T}^2}{m^2}),\nonumber
	\end{eqnarray}
	In the large mass approximation, it is instructive to use EFT to consider the new particles implication to $gg\rightarrow hh$ process.

If we now consider BSM modification to the di-Higgs production process, we have the parton cross section:
	\begin{eqnarray}\label{30}
	\lvert\M\lvert^2=\frac{\alpha_s^2G_F^2\hat{s}^2}{2^{17}\pi^2}\Lvert\sum_i\beta_i\big(g_{h}^{(i)} C_{\triangle}+(-g_h^{(i)} +g_{hh}^{(i)} )C_{\Box}\big)\Lvert^2
	\end{eqnarray}
	
	According to the definition given in Eq.~(\ref{30}), $\kappa_g=\sum_i \beta_ig_{h}^{(i)}/(\beta_t g_h^t)$ is the coupling of the Higgs to a pair of gluons, normalized by the SM induced one. Once the set of couplings, $\vec{\kappa}$, is introduced to parametrize deviations from the SM couplings of the Higgs bosons couplings to SM bosons and fermions~\cite{Heinemeyer:2013tqa}, one could compute the Higgs cross section obtain by the annihilation of a particle $i$ by $\kappa_i^2 = \sigma_i/\sigma_i^{SM}$. 
 
Therefore, using Eq.~(\ref{eqn:stopmassmatrix}) and~(\ref{eqn:stopmasses}), the couplings $g_h^{\tilde t}$ and $g_{hh}^{\tilde t}$ are given by :
 \begin{eqnarray}\label{10}
   g_{h}^{\tilde{t}}=\frac{\partial\log {(\det M_{\tilde{t}}^{2}(\upsilon)})}{\partial\log\upsilon} & = & 2m_t^2\frac{m_{\tilde{t}_1}^2+m_{\tilde{t}_2}^2-X_t^2}{m_{\tilde{t}_1}^2m_{\tilde{t}_2}^2}\\
   \label{11}
   g_{hh}^{\tilde{t}}=\frac{\partial^{2}\log {(\det M_{\tilde{t}}^{2}(\upsilon)})}{\partial(\log\upsilon)^{2}} & = & 2g_{h}^{\tilde{t}}-{g_{h}^{\tilde{t}}}^{2}+\frac{8m_t^4}{m_{\tilde{t}_1}^2m_{\tilde{t}_2}^2},
   \end{eqnarray}

 The coupling  of single Higgs boson production via gluon fusion process $gg\rightarrow h$, $\kappa_g$, is defined to be 1 in the SM based on the contribution form top and bottom quarks, and if we consider the contribution of top squarks running in the QCD loop, it would be given by :
 
 \begin{eqnarray}\label{kappag}
   \kappa_{g}=\left( 1+\frac{1}{8}g_{h}^{\tilde{t}}\right)\kappa_t
   \end{eqnarray}
 This could still be taken as a good approximation, and we use this equation to constrain the values of $\kappa_g$ we used in this paper.

\end{document}